\documentclass[11pt]{article}
\pdfoutput=1
\usepackage[margin=1.0in]{geometry}
\usepackage{amsmath,amssymb,graphicx}
\usepackage{hyperref}
\usepackage{slashed}

\usepackage[utf8]{inputenc}

\usepackage{pdflscape}

\usepackage{upgreek}
\usepackage{booktabs}

\usepackage{breqn}

\usepackage[font=small,labelfont=bf]{caption}
\usepackage{cite}
\usepackage{enumerate}

\usepackage[table]{xcolor}

\usepackage{makecell}
\usepackage{bm}
\definecolor{gray}{cmyk}{0,0,0,0.05}

\usepackage{multirow}
\usepackage{changepage}

\usepackage{array}
\newcolumntype{C}[1]{>{\centering\let\\\arraybackslash\hspace{0pt}}m{#1}}

\usepackage{tikz}
\usetikzlibrary{trees}
\usetikzlibrary{decorations.pathmorphing}
\usetikzlibrary{decorations.markings}
\tikzset{  
    fermion/.style={draw=black, postaction={decorate},
        decoration={markings,mark=at position .55 with {\arrow[draw=black]{>}}}},
    scalar/.style={draw=black, dashed,postaction={decorate},
        decoration={markings,mark=at position .55 with {\arrow[draw=black]{>}}}}
}

\newcommand{\iu}{{\mathrm i}}

\newcommand{\be}{\begin{equation}}
\newcommand{\ee}{\end{equation}}
\newcommand{\beq}{\begin{equation}}
\newcommand{\eeq}{\end{equation}}

\definecolor{gray}{cmyk}{0,0,0,0.05}
\newcolumntype{a}{>{\columncolor{gray}} l}

\newcommand{\sigmabar}{{\overline \sigma}}

\numberwithin{equation}{section}

\title{\bf  Interpreting the Electron EDM Constraint}
\author{Cari Cesarotti,$^{a}$ Qianshu Lu,$^{a}$ Yuichiro Nakai,$^{b}$ Aditya Parikh,$^{a}$ and Matthew Reece$^{a}$ \\
{\small $^{a}$ Department of Physics, Harvard University, Cambridge, MA, 02138}\\
{\small $^{b}$ Department of Physics and Astronomy, Rutgers University, Piscataway, NJ 08854}}

\begin{document}
\maketitle

\begin{abstract}
The ACME collaboration has recently announced a new constraint on the electron EDM, $|d_e| < 1.1 \times 10^{-29}\, e\, {\rm cm}$, from measurements of the ThO molecule. This is a powerful constraint on CP-violating new physics: even new physics generating the EDM at two loops is constrained at the multi-TeV scale. We interpret the bound in the context of different scenarios for new physics: a general order-of-magnitude analysis for both the electron EDM and the CP-odd electron-nucleon coupling; 1-loop SUSY, probing sleptons above 10 TeV; 2-loop SUSY, probing multi-TeV charginos or stops; and finally, new physics that generates the EDM via the charm quark or top quark Yukawa couplings. In the last scenario, new physics generates a ``QULE operator'' $(q_f \sigmabar^{\mu \nu}{\bar u}_f) \cdot (\ell {\sigmabar}_{\mu \nu} {\bar e})$, which in turn generates the EDM through RG evolution. If the QULE operator is generated at tree level, this corresponds to a previously studied leptoquark model. For the first time, we also classify scenarios in which the QULE operator is generated at one loop through a box diagram, which include (among others) SUSY and leptoquark models. The electron EDM bound is the leading constraint on a wide variety of theories of CP-violating new physics interacting with the Higgs boson or the top quark. We argue that any future nonzero measurement of an electron EDM will provide a strong motivation for constructing new colliders at the highest feasible energies.
\end{abstract}

\section{Introduction}

The ACME collaboration has used ThO molecules to constrain the electron electric dipole moment (EDM) to be \cite{Andreev:2018ayy}
\be
|d_e| < 1.1 \times 10^{-29}\, e\, {\rm cm}.   \label{eq:WOW}
\ee
This is about an order of magnitude improvement on the previous bound from ACME \cite{Baron:2013eja} and from studies of HfF$^+$ at JILA \cite{Cairncross:2017fip}. A nonzero electron EDM would establish physics beyond the Standard Model. The electron EDM violates CP (or equivalently, T) symmetry. In the Standard Model, this symmetry is violated by a handful of parameters: the CKM phase, which generates an electron EDM only at four loops with $|d_e| \sim 10^{-44}\, e\, {\rm cm}$ but also a CP-odd electron-nucleon interaction that can mimic an EDM of size $|d_e| \sim 10^{-38}\, e\, {\rm cm}$ \cite{Pospelov:2013sca} (see \cite{Hoogeveen:1990cb, Pospelov:1991zt} for earlier work); the strong phase $\bar \theta$, which generates an electron EDM $|d_e| \lesssim 10^{-37}\, e\, {\rm cm}$ \cite{Choi:1990cn, Ghosh:2017uqq}; and phases associated with the lepton sector, which give contributions at two loops suppressed by neutrino masses \cite{Ng:1995cs} with an expectation that $|d_e| \lesssim 10^{-43}\, e\, {\rm cm}$ or, in the presence of severe fine-tuning, at most $|d_e| \lesssim 10^{-33}\, e\, {\rm cm}$ \cite{Archambault:2004td}.  As a result, it is of great interest to continue searching for a smaller electron EDM consistent with \eqref{eq:WOW} but inconsistent with the Standard Model.

The recent progress in EDM searches comes at a key time in the field of particle physics.  The discovery of the Higgs boson at the LHC filled in the last missing piece of the Standard Model. While there are many motivations for searching for physics beyond the Standard Model, three of the most important are the matter-antimatter asymmetry of our universe, the existence of dark matter, and the fine-tuning puzzle of the Higgs boson mass. The matter-antimatter asymmetry clearly indicates a need for new CP-violating physics, which could first be detected through its indirect effect on the electron EDM. As we will discuss below, EDMs also have interesting connections with WIMP dark matter (in specific models) and with the fine-tuning problem. 

The possibility of testing heavy new physics through electric dipole moment measurements has been studied extensively; reviews include \cite{Pospelov:2005pr, Engel:2013lsa, Safronova:2017xyt, Chupp:2017rkp}. Here we attempt to briefly summarize some of the important history of the topic, with apologies for inevitable omissions. Some early theoretical studies of lepton EDMs appeared already in the 1970s \cite{Pais:1974xf, Donoghue:1977bw}. Many of the early studies of CP violation in supersymmetric theories focused on the neutron EDM \cite{Ellis:1982tk, Chia:1982gp, Polchinski:1983zd}, but studies of the electron EDM in supersymmetry commenced \cite{delAguila:1983dfr} shortly after a suggestion of Gavela and Georgi that lepton EDMs could be effective probes of new physics \cite{GavelaLegazpi:1982ud}. Subsequently, a variety of additional sources of EDMs were studied, such as 3-gluon operators \cite{Weinberg:1989dx} or two-loop diagrams mediated by electroweak bosons \cite{Barr:1990vd, Leigh:1990kf}. A variety of new physics scenarios have been shown to predict interesting EDMs, including: stops in SUSY \cite{Chang:1998uc}; electroweakinos in SUSY \cite{Li:2008kz} and specifically split SUSY \cite{ArkaniHamed:2004yi, Giudice:2005rz}; two Higgs doublet models \cite{Leigh:1990kf,Chang:1990sf,Jung:2013hka, Abe:2013qla}; SUSY beyond the MSSM \cite{Blum:2010by, Altmannshofer:2011rm}; and fermionic top partners \cite{Panico:2017vlk}. 

Our goal in this paper is to give a brief survey of how theories of new physics are constrained by the ACME result, including a range of novel possibilities where an EDM is mediated by the charm or top quark. We begin in \S\ref{sec:bigpicture} by giving a general argument, based on effective field theory, for the range of mass scales that are probed by the EDM. In scenarios with two-loop EDMs where the electron Yukawa coupling appears explicitly in the new physics couplings---which includes many SUSY scenarios---the ACME constraint probes masses of a few TeV. Other scenarios, where loop effects generate both the EDM and the electron Yukawa coupling, potentially probe scales of hundreds of TeV. We also discuss the case where the dominant effect on ThO is not the electron EDM at all but the CP-odd electron-nucleon coupling (as discussed in e.g.~\cite{Lebedev:2002ne,Demir:2003js, Pospelov:2005pr, Jung:2013mg, Chupp:2014gka}). Next we turn to a discussion of EDM constraints on supersymmetric scenarios: one-loop SUSY in \S\ref{sec:oneloopSUSY}; two-loop split SUSY in \S\ref{sec:twoloopsplit}; and two-loop natural SUSY in \S\ref{sec:twoloopnatural}. Our calculations in the two loop cases follow \cite{Nakai:2016atk}, which can be consulted for more details. Our discussion of split SUSY includes a comparison of the reach of EDMs and of recent dark matter direct detection results from Xenon1T \cite{Aprile:2018dbl}. Most of our discussion of SUSY draws heavily on earlier literature, but updates it in light of the new experimental result. In the context of one-loop effects, if sleptons and squarks are at a comparable mass scale then we show that there is an interesting complementarity between the requirement of consistency with a 125 GeV Higgs (which prevents making the scalar mass too large, for any given $\tan \beta$) and the EDM constraint (which prevents the scalars from being too light, for a given CP-violating phase). This interesting qualitative point is shown in Fig.~\ref{fig:winohiggsinoslepton}, which provides a novel way to visualize how EDM experiments are encroaching on the viable SUSY parameter space. In \S\ref{sec:QULE} we discuss the possibility that the EDM is induced by the QULE operator $(q_f \sigmabar^{\mu \nu}{\bar u}_f) \cdot (\ell {\sigmabar}_{\mu \nu} {\bar e})$. In this case new physics need not couple to the Higgs boson at all to generate an EDM. Instead, new physics couples quarks and leptons, and then the quark Yukawa coupling supplies the necessary interaction with the Higgs. The most plausible version of this scenario has the top quark inducing the EDM, though the charm quark could also play this role. (If the up quark is the leading coupling, then the CP-odd electron-nucleon coupling plays a more important role in the ThO measurement than the electron EDM itself.) The QULE operator could be induced by scalar leptoquark exchange at tree level, as previously discussed in \cite{Arnold:2013cva, Fuyuto:2018scm,Dekens:2018bci}. It could also arise from a box diagram, a case that we discuss for the first time. We classify a number of possibilities for the quantum numbers of the particles appearing in the loop, which could have a variety of distinctive collider signals. In some of these models, the QULE-generated EDM is the leading contribution, as couplings allowing for other EDM contributions (e.g.~of Barr-Zee type) are absent. We conclude in \S\ref{sec:conclusions} with a few remarks on the implications of future improvements in EDM searches. If no EDM is detected, then the CP-violating phases associated with any new physics near the TeV scale will be constrained to be very small. We believe that it is timely to further investigate how naturally small CP-violating phases might be explained. Conversely, if a nonzero EDM is detected, then either it arises from TeV-scale particles that may be detected at future colliders, or from even heavier particles, out of reach of currently feasible experiments. We argue that in the latter case, these new particles would lead to a very concrete form of the hierarchy problem, motivating the construction of new colliders even if the particles directly contributing to the EDM are out of reach.

%%%%%%%%%%%%%%%%%%%%%%%%%%%%%%%%%%%%%%%%%%%%%%%
\section{Interpreting the EDM constraint: the big picture}
\label{sec:bigpicture}

In this section, we present a general argument for the range of mass scales probed by the EDM experiments. The cases where the dominant effect on ThO comes from the CP-odd electron-nucleon coupling as well as the electron EDM are discussed.

\subsection{The electron EDM}\label{edmgeneral}

Dipole moment interactions flip the chirality of a fermion. In the Standard Model, since left- and right-handed fermion fields carry different charges, gauge invariance requires that the EDM is a dimension-six operator involving the Higgs boson: either $\iu h^\dagger \ell \sigmabar^{\mu \nu} {\bar e} B_{\mu \nu}$ or $\iu h^\dagger \ell \sigma^i \sigmabar^{\mu \nu} {\bar e} W^i_{\mu \nu}$. Hence, an EDM generated by new physics at a mass scale $M$ is always proportional to $v/M^2$ times a product of couplings and loop factors. Physics that produces the EDM operators can also produce corrections to the electron Yukawa interaction $h^\dagger \ell {\bar e}$, simply by removing the gauge interaction vertex from the Feynman diagrams that appear in the EDM calculation. As a result, we expect the size of the EDM to be bounded in terms of the size of the electron Yukawa in typical scenarios without fine-tuning.

We can consider three scenarios for how to treat the relationship between the EDM and the electron Yukawa:
\begin{itemize}
\item {\bf Spurion approach.} Here we assume that the couplings generating the EDM are directly proportional to $y_e$. If the EDM is generated at $k$ loops, we expect:
\be
d_e \sim \delta_{\rm CPV} \left(\frac{y^2}{16\pi^2}\right)^k \frac{m_e}{M^2},
\ee
with $y$ standing in for whatever coupling arises in the loop, generally presumed to be an order-one number, and $\delta_{\rm CPV}$ the size of the CP-violating phase.
\item {\bf Radiative stability approach.} Here we make the weaker assumption that the interactions generating the EDM make no more than an order-one change to the size of the electron Yukawa coupling. This could be the case, for example, if the electron Yukawa coupling is radiatively generated by the same interactions. We have:
\be
d_e \sim \delta_{\rm CPV} \frac{m_e}{M^2}.
\ee
In this case the estimate matches the 0-loop spurion estimate, as we assume that the same loop factors are shared by $y_e$ and $d_e$.
\item {\bf Tuning approach.} In this case, we allow for the interactions generating the EDM to generate a contribution to the electron Yukawa much larger than the Yukawa itself, so that the final Yukawa is tuned to be small via a cancelation. This is the least aesthetically appealing case, but is a logical possibility. The EDM can arise from a $k$-loop diagram containing $2k+1$ Yukawa couplings, each in principle as large as $y \lesssim 4\pi$, the value estimated by Naive Dimensional Analysis (NDA). Hence we estimate a maximum EDM consistent with NDA:
\be
d_e \sim \delta_{\rm CPV} \, y \left(\frac{y^2}{16\pi^2}\right)^k  \frac{v}{M^2} \lesssim
\delta_{\rm CPV}   \frac{4 \pi  v}{M^2}.
\ee
The tuning approach allows for the largest mass scale for new physics.
\end{itemize}

Following these simple estimates and taking, for concreteness, $y = g$ with $g \approx 0.65$ the weak coupling constant, we obtain the following rough estimates of the mass scales of new physics probed by the EDM measurement \eqref{eq:WOW}: 
{\Large
{ \setlength{\tabcolsep}{1.5em}
\begin{center}
Mass Scale of New Physics Necessary for $|d_e| \lesssim 1.1 \times 10^{-29}\, e\, {\rm cm}$
\begin{tabular}{ a  c  c c}
\rowcolor{gray} & {\bf 0-loop} & {\bf 1-loop} & {\bf 2-loop} \\
{\bf Spurion} & 1000 TeV & 50 TeV & 3 TeV \\
{\bf Radiative} & 1000 TeV & 1000 TeV & 1000 TeV \\
{\bf Tuned} & $2 \times 10^9$ TeV & $2 \times 10^9$ TeV & $2 \times 10^9$ TeV 
\end{tabular}
\end{center} }
}
We see that with fine tuning to cancel large corrections to the electron Yukawa, the EDM measurement can in principle probe physics far above that being studied at colliders. However, for more theoretically plausible models, the mass scale probed is below 1000 TeV and, in a wide range of models that lead to EDMs at two loops, is of order 1 TeV.

\begin{figure}[h]
  \centering
  \includegraphics[width=1.0\textwidth]{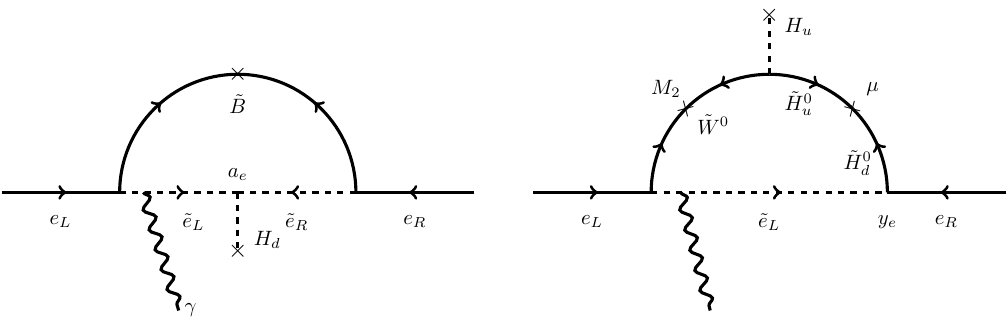}
  \caption{One-loop EDMs in supersymmetric theories.}
  \label{fig:oneloopEDM}
\end{figure}

\begin{figure}[h]
\begin{center}
  \includegraphics[width=0.5\textwidth]{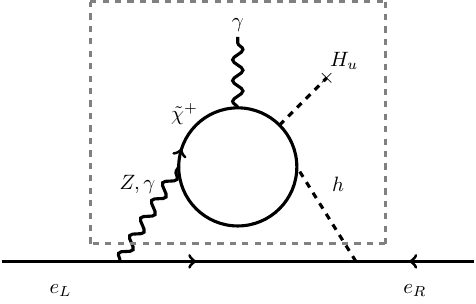}
\end{center}
\caption{Two-loop EDMs in supersymmetric theories. The one-loop diagram in the dashed box is a CP-violating analogue of familiar ``electroweak precision'' corrections.}
\label{fig:twoloopEDM}
\end{figure}

In later sections, we will see the simple estimates in the table substantiated in a variety of concrete models of new physics, but
let us briefly summarize the models we expect in each category of the table.
First, we look at tree-level contributions to the electron EDM.
If the lepton sector couples to some new strongly-coupled sector with CP violation, we do not expect any loop suppression factor
in the expression of the EDM.
In the case where the physics giving the electron chirality flip is still proportional to the electron Yukawa,
this corresponds to the tree-level spurion approach.
On the other hand, in the scenario of partial compositeness where the lepton sector linearly couples
to a new strongly-coupled sector giving a composite Higgs
\cite{Kaplan:1991dc},
the electron EDM is generated by the same interactions realizing the electron Yukawa coupling,
which is the tree-level radiative stability approach.

An example of a one-loop EDM in SUSY theories is shown in Figure~\ref{fig:oneloopEDM}. The right diagram is explicitly proportional to the electron Yukawa while the left diagram is not. However, depending on the proportionality of the $A$-term to $y_e$, this contribution is classified into the spurion scenario or the other scenarios. These diagrams illustrate an important general point: one-loop diagrams for the EDM will generally contain {\em some new particle with lepton quantum numbers}, like the electron superpartner appearing in this diagram. If all new particles with lepton quantum numbers are heavy, there may be no important one-loop diagrams contributing to the EDM, and the most important contributions may arise at two loops.

A two-loop electron EDM can arise from the Barr-Zee type diagrams \cite{Barr:1990vd}, from similar diagrams induced by the $W$ boson EDM \cite{Atwood:1990cm}, or from rainbow diagrams \cite{Yamanaka:2012ia}. They are all proportional to the electron Yukawa coupling and classified into the spurion approach. The two-loop EDM can also arise from RGE running induced by $q{\bar u}\ell {\bar e}$-type operators arising from a one-loop box diagram \cite{Alonso:2013hga} which provides an example of the radiative stability approach as we will study in detail later. An example of a two-loop EDM in SUSY theories is shown in Figure~\ref{fig:twoloopEDM}.

\subsection{An alternative: the CP-odd electron-nucleon coupling}
\label{subsec:CScontrib}

We should consider the possibility that the electron EDM inferred from ThO is not really the electron EDM at all, but instead evidence of a CP-odd electron-nucleon interaction $-\iu C_S {\bar e} \gamma_5 e {\bar N} N$. Indeed, in the Standard Model this is expected to be a larger effect \cite{Pospelov:2013sca}, though we will argue that the opposite is true for many models of new physics. The coupling $C_S$ contributes to the effective EDM as \cite{Chupp:2014gka}
\begin{align}
d_{\rm ThO} &\approx d_e + k C_S, \nonumber \\
k &\approx 1.6 \times 10^{-15}\, {\rm GeV}^2\, e\, {\rm cm}.  
     \label{eq:dThO}
\end{align}
The size of $k$ depends in a somewhat nontrivial way on factors including the ratio of atomic and nuclear radii and the value of $Z\alpha$ for the atoms involved; we refer readers to the appendix of \cite{dzuba2011relations} for details. The microscopic origin of such a four-fermion interaction is in similar interactions with quarks in place of nucleons:
\begin{equation}
\begin{split}
\mathcal{L}_{\rm Four-Fermi} \supset \sum_{q}  C_{q e} \left( \bar{q} q \right) \left( \bar{e} \, {\rm i} \gamma_5 e \right) , \label{fourfermiint}
\end{split}
\end{equation}
where $q$ denotes any flavor of quarks.
These four-fermion interactions lead to
\begin{equation}
\begin{split}
C_S &\approx C_{de} \langle N | {\bar d}d | N \rangle + C_{se} \langle N | {\bar s}s | N \rangle + C_{be} \langle N | {\bar b}b | N \rangle
\\[1ex]
&\quad+C_{ue} \langle N | {\bar u}u | N \rangle + C_{ce} \langle N | {\bar c}c | N \rangle + C_{te} \langle N | {\bar t}t | N \rangle \\[1ex]
&\approx C_{de} \frac{29\, {\rm MeV}}{m_d} + C_{se} \frac{49\, {\rm MeV}}{m_s}  + C_{be} \frac{74\, {\rm MeV}}{m_b} \\[1ex]
&\quad+C_{ue} \frac{16\, {\rm MeV}}{m_u} + C_{ce} \frac{76\, {\rm MeV}}{m_c}  + C_{te} \frac{77\, {\rm MeV}}{m_t} .
\label{CSexpression}
\end{split}
\end{equation}
Here, we have used the matrix elements
\cite{Ellis:2008zy,Junnarkar:2013ac},
\begin{equation}
\begin{split}
&(m_u+m_d) \langle  N | \bar{u}u + \bar{d}d | N \rangle \simeq 90 \, {\rm MeV} ,
\qquad \langle  N | \bar{u}u - \bar{d}d | N \rangle \simeq 0 , \\[1ex]
&m_s \langle  N | \bar{s}s | N \rangle \simeq 49 \, {\rm MeV} , \qquad m_b \langle  N | \bar{b}b | N \rangle \simeq 74 \, {\rm MeV} , \\[1ex]
&m_c \langle  N | \bar{c}c | N \rangle \simeq  76 \, {\rm MeV} , \qquad m_t \langle  N | \bar{t}t | N \rangle \simeq 77 \, {\rm MeV} .
\end{split}
\end{equation}
and $m_u / m_d =0.55$.

As in the case of the electron EDM, physics generating the $C_S$ coupling can also produce corrections to the electron mass by connecting two quark legs of \eqref{fourfermiint} with an insertion of the quark mass, and we can consider three scenarios in the relationship between the $C_S$ coupling
and the Yukawa couplings:
\begin{itemize}
\item {\bf Spurion approach.} We assume that the couplings generating the four-fermion interactions are directly proportional to the electron and quark Yukawa couplings.
If the four-fermion interactions are generated at $k$ loops, we expect:
\begin{equation}
\begin{split}
C_{qe} \sim \delta_{\rm CPV}  \left(\frac{y^2}{16\pi^2}\right)^k \frac{m_q m_{e}}{v^2M^2}. 
\end{split}
\end{equation}
In this scenario, the quark mass suppression in \eqref{CSexpression} is cancelled by the quark mass dependence in $C_{qe}$.
If we take this ansatz for all of the quarks, the top quark gives the most important contribution.

\item {\bf Radiative stability approach.} As in the case of the electron EDM, we make the weaker assumption that the generated four-fermion interactions do not lead to more than an order-one change to the size of the electron Yukawa coupling, which gives a constraint on the size of the coupling in an underlying theory to generate the four-fermion interactions.
Then, we expect:
\begin{equation}
\begin{split}
C_{qe} \sim  \delta_{\rm CPV} \frac{16\pi^2 m_e}{m_q} \frac{1}{M^2} .
\end{split}
\end{equation}
Due to the quark mass suppression in this expression as well as in \eqref{CSexpression}, the up quark gives the most important contribution (assuming that we take this ansatz for all of the quarks).
Since $m_q > m_e$, the requirement that the four-fermion interactions do not lead to more than
an order-one change to the size of quark Yukawa couplings does not lead to a further constraint on the size of $C_{qe}$.

Notice that because the operator breaks both quark and lepton chiral symmetries, if it has a large coefficient one can think of it as leaving invariant only a combined chiral rotation of both quarks and leptons.

\item {\bf Tuning approach.} 
In this scenario, we allow for underlying interactions generating the four-fermion interactions to generate a contribution to the electron Yukawa much larger than the correct size, so that the final electron Yukawa is tuned to be small.
If the four-fermion interactions are generated at $k$ loops, we expect from NDA:
\begin{equation}
\begin{split}
C_{qe} \sim \delta_{\rm CPV}  \, y^2 \left( \frac{y^{2}}{16\pi^2} \right)^k \frac{1}{M^2}
&\lesssim \delta_{\rm CPV}  \frac{16\pi^2}{M^2}. 
\end{split}
\end{equation}
The quark mass suppression in \eqref{CSexpression} makes the top quark contribution naively very small and the up quark gives the most important contribution.

\end{itemize}

Following these simple estimates and taking $y = g$, we obtain the following rough estimates of the mass scales of new physics probed by the EDM measurement (the parenthesis denotes the dominant contribution in each category):
{%\Large
{ \setlength{\tabcolsep}{1.5em}
\begin{center}
Mass Scale of New Physics Necessary for $|d_{\rm ThO}| \lesssim  1.1 \times 10^{-29}\, e\, {\rm cm}$
\begin{tabular}{ a  c  c c}
\rowcolor{gray} & {\bf 0-loop} & {\bf 1-loop} & {\bf 2-loop} \\
{\bf Spurion} &  300 GeV ($C_{te}$) & 20 GeV ($C_{te}$) & 0.8 GeV ($C_{te}$) \\
{\bf Radiative} & $1 \times 10^{5}$ TeV ($C_{ue}$) & $1 \times 10^{5}$ TeV ($C_{ue}$) & $1 \times 10^{5}$ TeV ($C_{ue}$) \\
{\bf Tuned} & $4 \times 10^{5}$ TeV ($C_{ue}$) & $4 \times 10^{5}$ TeV ($C_{ue}$) & $4 \times 10^{5}$ TeV ($C_{ue}$)
\end{tabular}
\end{center} }
}

In listing the dominant contribution we assume the same ansatz applies for all quarks; of course, a more general flavor structure, including the possibility of off-diagonal couplings, is also possible, but the simple ansatz gives a qualitative sense of the range of mass scales of interest.
The tuned and radiative stability approaches probe large energy scales. On the other hand, in the spurion approach, the mass scale is below 1 TeV even in the 0-loop case and has been already explored at colliders.

%%%%%%%%%%%%%%%%%%%%%%%%%%%%%%%%%%%%%%%%%%%%%%%%%%%%%%%%%

\section{The EDM constraint on one-loop SUSY}
\label{sec:oneloopSUSY}

\begin{figure}[!t]
  \centering
  \includegraphics[width=0.43\textwidth]{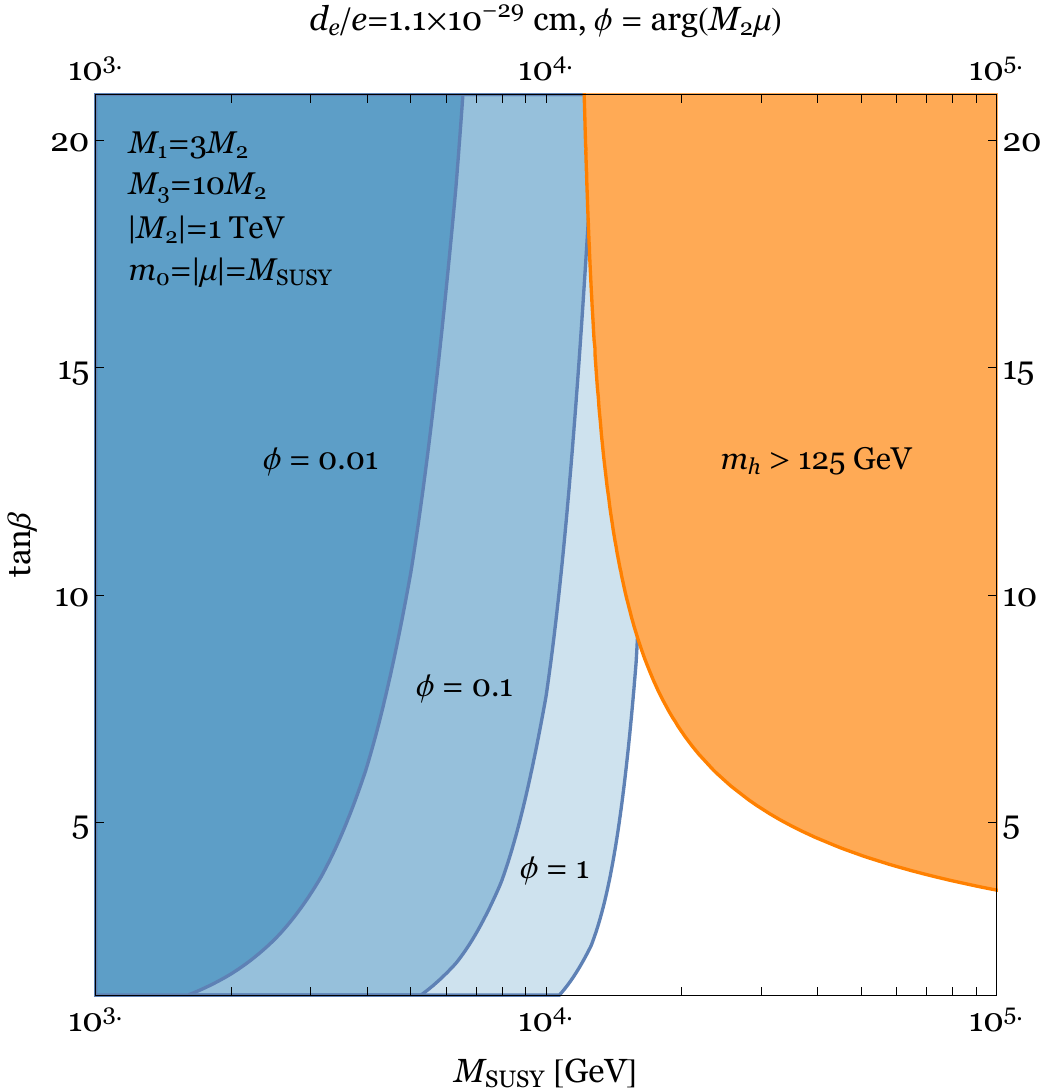}  
  \includegraphics[width=0.43\textwidth]{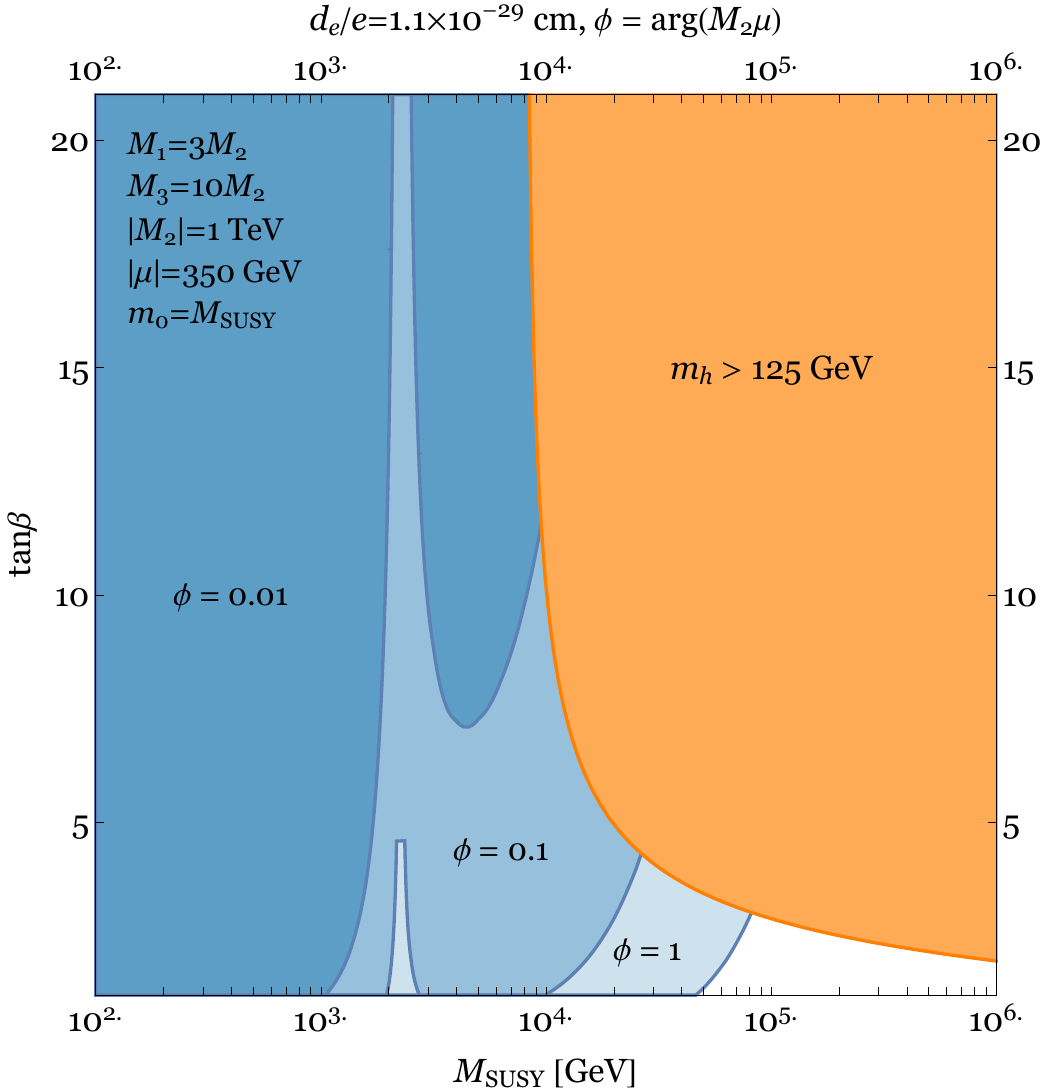} \\
  \includegraphics[width=0.43\textwidth]{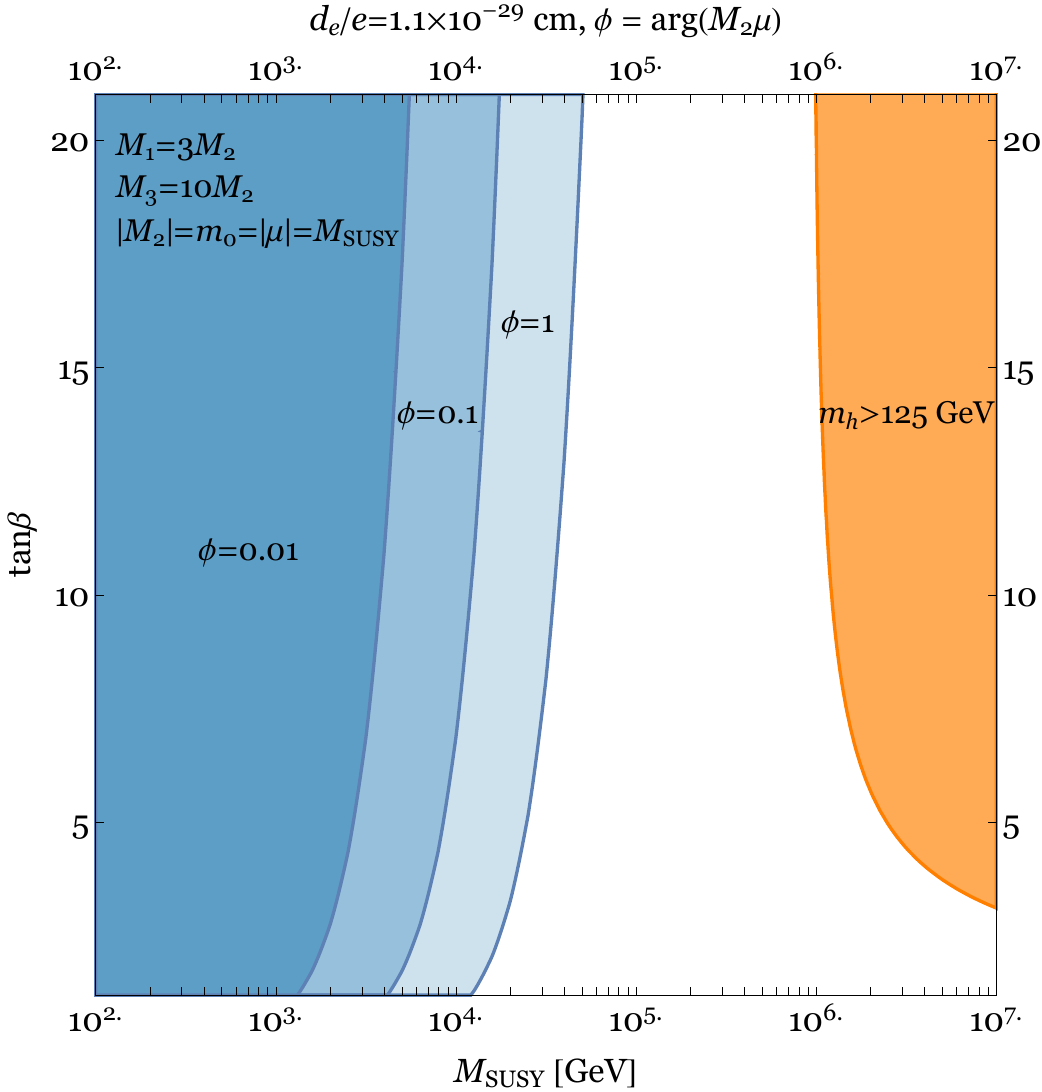}
  \caption{The ACME II constraint on $M_{\rm SUSY}$ and $\tan \beta$. The electron EDM is generated by the right diagram of Fig.~\ref{fig:oneloopEDM}. The orange region is excluded from $m_h > 125 \, \rm GeV$.
  The upper left and right panels correspond to the case of split SUSY,  $M_{1,2,3} \ll m_0 = M_{\rm SUSY}$.
  We take $|\mu| = M_{\rm SUSY}$ and $|\mu| = 350 \, \rm GeV$
  in the left and right panels respectively. The lower panel corresponds to the case of high-scale SUSY,
  $M_{1,2,3} \sim m_0 = M_{\rm SUSY}$. In all cases, we assume
  a gaugino mass ratio, $M_1: M_2: M_3 = 3: 1: 10$. In each panel, we plot three cases of the phase
  $\phi \equiv \arg (M_2 \mu) = 1, 0.1, 0.01$.}
  \label{fig:winohiggsinoslepton}
\end{figure}

In this section we discuss constraints on supersymmetry arising from 1-loop EDMs. The relevant formulas are well-known in the literature (e.g.~\cite{Ibrahim:2007fb}), but it is useful to update the bounds in light of new data. Furthermore, by comparing the parameter space ruled out by EDMs with the parameter space in which the MSSM cannot accommodate the measured Higgs mass, we provide a novel visualization of the power of EDM searches (see Fig.~\ref{fig:winohiggsinoslepton}).

We showed examples of one-loop SUSY EDMs above in Fig.~\ref{fig:oneloopEDM}. To unpack the diagrams a bit more: the electron splits into a virtual pair of its superpartner (the selectron) and a neutralino (the superpartner of the photon, $Z$, or Higgs boson). The diagram at right contains a selectron--electron--Higgsino interaction, which depends on the electron Yukawa coupling $y_e = m_e/v$. Then, it is proportional to $m_e$. The diagram at {\em left}, on the other hand, transforms the left-handed selectron to the right-handed selectron using the $A$-term trilinear coupling, $a_e H_d {\tilde e}_L {\tilde e}_R$. In a general supersymmetric theory, $a_e$ is formally independent of the Yukawa coupling $y_e$, although in many models they are proportional: $a_e \approx y_e M_{\rm SUSY}$, where $M_{\rm SUSY}$ is some measure of the SUSY-breaking scale. Since attempting to break this proportionality would lead to large corrections to $m_e$, it is reasonable to assume the proportionality. In this section, we concentrate on flavor-diagonal contributions to the EDM, which exist even in the absence of flavor violation in soft scalar masses generating dangerous FCNCs. If there are large off-diagonal scalar mass terms, different diagrams with insertion of scalar mass mixing become important \cite{McKeen:2013dma,Altmannshofer:2013lfa}.

In the diagram at left, the invariant phase that would contribute to CP violation is $\arg(a_e^* M_{1,2})$. In many particular models of SUSY breaking, like gauge mediation (for reviews, \cite{Giudice:1998bp,Kitano:2010fa}), this CP phase is zero, and the contribution is absent. In more general models, like gravity mediation, it is unclear whether we should expect this phase to be small. The diagram at right is sensitive to the phase $\arg(\mu M_2 b_\mu^*)$.\footnote{Much of the literature performs a field redefinition to remove the phase of $b_\mu$ and refers to this simply as $\arg(\mu M_2)$, which we will sometimes write below.} Generation of $\mu$, the Higgsino mass parameter, is typically one of the thorniest problems in building a supersymmetric model, and it seems very plausible that it could have a CP phase different from other SUSY-breaking parameters.

Let us summarize a general one-loop formula of the fermion EDM induced by a fermion $\psi_i$ with mass $m_i$
and electric charge $Q_i$
and a complex scalar $\phi_j$ with mass $m_j$ and electric charge $Q_j$. Their interactions with the SM fermion $f$ are: 
\begin{equation}
\mathcal{L}_{\psi f \phi} = L_{ij} \left( \bar{\psi}_i P_L f\right) {\phi}^*_j + R_{ij} \left( \bar{\psi}_i P_R f\right) {\phi}^*_j + {\rm h.c.} ,
\end{equation}
where $P_{R, L} = \frac{1}{2}\left(1 \pm\gamma^5 \right)$ and $L_{ij}$ and $R_{ij}$ are coupling constants.
With these interaction terms, the one-loop calculation gives \cite{Ibrahim:2007fb}
\begin{equation}
    \begin{split}
\frac{d_f}{e} &=  \left(\frac{ m_i}{16\pi^2m_j^2}\right) {\rm Im}{[(R_{ij})^*L_{ij}]}\left[Q_i \frac{1}{2(1-r)^2}\left(\frac{2\log r}{1-r}+3-r\right) + Q_j \left(\frac{1+r+\frac{2r \log(r)}{1-r}}{2(1-r)^2} \right) \right] . \label{oneloopgeneral}
    \end{split}
\end{equation}
Here, $r = m_i^2/m_j^2$.

\begin{figure}[!t]
  \centering
  \includegraphics[width=0.5\textwidth]{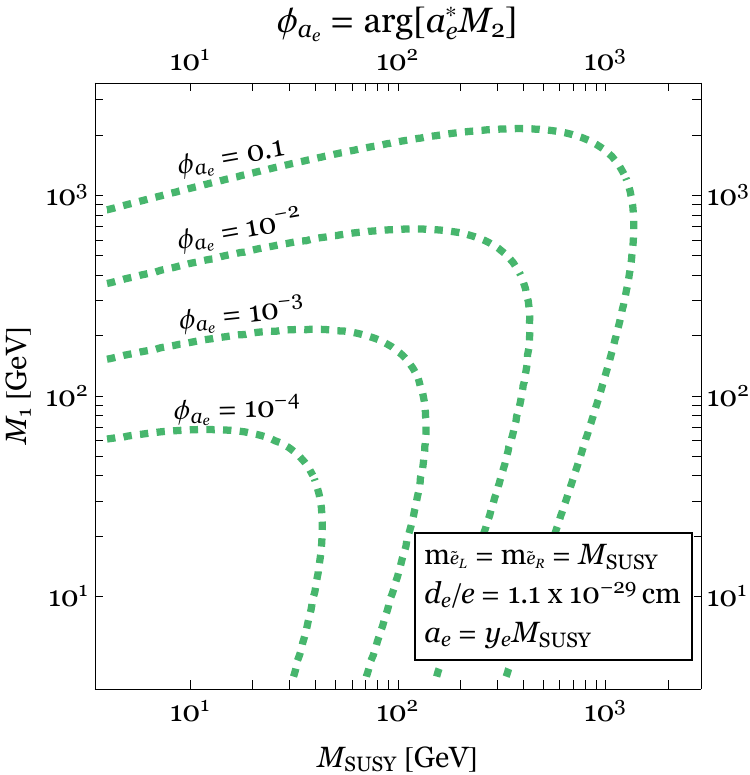}
  \caption{The ACME II constraint on the selectron mass scale $M_{\rm SUSY}$ and the Bino mass $M_1$.
  The left region of each contour is excluded.
  We plot four cases of the phase $\phi \equiv \arg [a_e^\ast M_2] = 0.1, 10^{-2}, 10^{-3}, 10^{-4}$. 
  We assume the left-handed and right-handed selectron soft masses are the same, $m_{\tilde{e}_L} = m_{\tilde{e}_R} = M_{\rm SUSY}$.}
  \label{fig:binoselectron}
\end{figure}

We now look at the contribution to the electron EDM from the diagram at the left of Fig.~\ref{fig:oneloopEDM}.
The mass terms of the selectrons and the Bino and their interaction terms (in two-component spinor notation) are
\begin{equation}
    \begin{split}
\mathcal{L} &\supset -m_{\tilde{e}_L}^2|\tilde{e}_L|^2-m_{\tilde{e}_R}^2|\tilde{{e}}_R|^2
-\left(\frac{1}{2}M_1\tilde{B}\tilde{B}+{\rm h.c.}\right)\\[1ex]
&-  a_e H_d \, \tilde{e}_L\tilde{{e}}_R^\ast 
-g'Y_L\tilde{B} e_L\tilde{e}_L^{\ast} -g'Y_{R} \tilde{B} {e}_{R}^\dagger \tilde{{e}}_R+ \rm h.c. ,
    \end{split}
\end{equation}
where $g'$ is the $U(1)_Y$ coupling constant and
$Y_L$ and $Y_R$ are the hypercharges of the left-handed and right-handed electrons respectively.
With a nonzero vev of $H_d$, the $A$-term gives an off-diagonal component in the selectron mass matrix.
We first diagonalize the mass matrix and rewrite the interactions in terms of the mass-eigenstate basis.
Then, using the above general formula \eqref{oneloopgeneral}, we can obtain the expression for the electron EDM $d_e$.
There are also contributions from the other neutralinos in the similar loops,
but they are subdominant when the gaugino masses are large enough compared to the electroweak breaking scale and ignored.
Figure~\ref{fig:binoselectron} shows the ACME II constraint on the selectron mass scale $M_{\rm SUSY}$ and the Bino mass $M_1$.
The left region of each contour is excluded.
We plot four cases of the phase $\phi \equiv \arg [a_e^\ast M_2] = 0.1, 10^{-2}, 10^{-3}, 10^{-4}$. 
We assume that the left-handed and right-handed selectron soft masses are the same, $m_{\tilde{e}_L} = m_{\tilde{e}_R} = M_{\rm SUSY}$.
Since the diagram contains an insertion of the vev $v_d$ and
$a_e = y_e M_{\rm SUSY}$ is assumed, the contribution to the EDM is proportional to $a_e v_d = m_e M_{\rm SUSY}$
and does not depend on $\tan \beta$.\footnote{Choosing $a/y$ of the same order as the soft masses can, in the stop sector, lead to color- and charge-breaking vacua \cite{Claudson:1983et,Casas:1996wj}. For $a_t/y_t = m_{{\tilde Q}_3} = m_{{\tilde u}_3} = M_{\rm SUSY}$ there is no vacuum stability problem, but for $a_t/y_t$ an order-one factor larger there could be \cite{Chowdhury:2013dka,Blinov:2013fta}.}
Taking into account the size of the $U(1)_Y$ coupling $(g'/g)^2 \approx 0.3$, the obtained lower bound on the mass is consistent with our general argument presented in section~\ref{edmgeneral}.

We next consider the diagram at the right of Fig.~\ref{fig:oneloopEDM}.
As above, the contribution to the electron EDM can be calculated by applying the general formula \eqref{oneloopgeneral}.
The relevant interaction terms should be given in terms of the basis of the neutralino and chargino eigenstates.
With the resulting expression of the electron EDM, the ACME II gives a direct constraint on slepton masses.
However, most scenarios of SUSY breaking, including gauge mediation and anomaly mediation \cite{Randall:1998uk,Giudice:1998xp}, generate squark and slepton soft masses at the same order of magnitude.
In addition, a large splitting between slepton and squark masses inside each generation induces a large one-loop effect of the hypercharge $D$-term, which may drive light scalars tachyonic.
Then, it is reasonable to assume squark and slepton masses at the same order and the null result of the EDM experiment
implies a lower bound on
squark masses as well as slepton masses.
Since top/stop loops give a significant radiative correction to the Higgs mass,
too-large stop masses may lead to a Higgs mass larger than $125 \, \rm GeV$,
which sets an upper bound on the mass scale of squarks and sleptons.

In Fig.~\ref{fig:winohiggsinoslepton}, we show constraints on the parameter space for EDMs from mixed electroweakinos and left-handed sleptons.
We compute the Higgs mass using SusyHD \cite{Vega:2015fna} assuming universal scalar masses $m_0$. 
The orange region is excluded from $m_h > 125 \, \rm GeV$.
The upper left and right panels correspond to the case of split SUSY,  $M_{1,2,3} \ll m_0 = M_{\rm SUSY}$.
We take $|\mu| = M_{\rm SUSY}$ and $|\mu| = 350 \, \rm GeV$
in the left and right panels respectively. The lower panel corresponds to the case of high-scale SUSY,
$M_{1,2,3} \sim m_0 = M_{\rm SUSY}$. In all cases, we assume
a gaugino mass ratio, $M_1: M_2: M_3 = 3: 1: 10$. In each panel, we plot three cases of the phase
$\phi \equiv \arg (M_2 \mu) = 1, 0.1, 0.01$.
In the upper two cases of split-SUSY, the EDM bound has already hit the excluded region of a too-large Higgs mass (provided $\tan \beta$ is not too small)
while there is still room between the EDM bound and the Higgs mass bound in the case of high-scale SUSY.

%%%%%%%%%%%%%%%%%%%%%%%%%%%%%%%%%%%%%%%%%%%%%
\section{The EDM constraint on two-loop split SUSY}
\label{sec:twoloopsplit}

In this and the following section, we will examine how the new experimental result allows us to update earlier conclusions about EDM constraints from 2-loop SUSY contributions. In split supersymmetry, if we decouple the squarks, sleptons, and heavy Higgs bosons (working at relatively low $\tan \beta$), the dominant EDMs will arise from loops of charginos and neutralinos \cite{ArkaniHamed:2004yi, Giudice:2005rz}. These are Barr-Zee type diagrams with an inner loop connected to the electron with $\gamma h$, $Z h$, or $WW$ propagators. The dominant diagram is $\gamma h$; $WW$ is not negligible, but $Zh$ is subleading since $\frac{1}{4} - \sin^2 \theta_W$ happens to be small. If we integrate out charginos at one loop we obtain the effective operator
\be
\frac{e^2}{16\pi^2} (\arg \det {\cal M}_{\widetilde C}) F_{\mu \nu} {\widetilde F}^{\mu \nu} = \frac{e^2}{8\pi^2} \frac{{\rm Im}(g^2 M_2 \mu H_u \cdot H_d)}{\left|M_2 \mu - g^2 H_u \cdot H_d\right|^2} F_{\mu \nu} {\widetilde F}^{\mu \nu}
\ee
where ${\cal M}_{\widetilde C}$ is the chargino mass matrix. This operator mixes into the EDM at one loop, allowing us to easily understand the leading-log contribution to the two-loop calculation. Because the numerator involves $H_u \cdot H_d$, the EDM becomes smaller at large $\tan \beta$ when the light Higgs boson has little overlap with $H_d$.

\begin{figure}[h]
  \centering
  \includegraphics[width=1.0\textwidth]{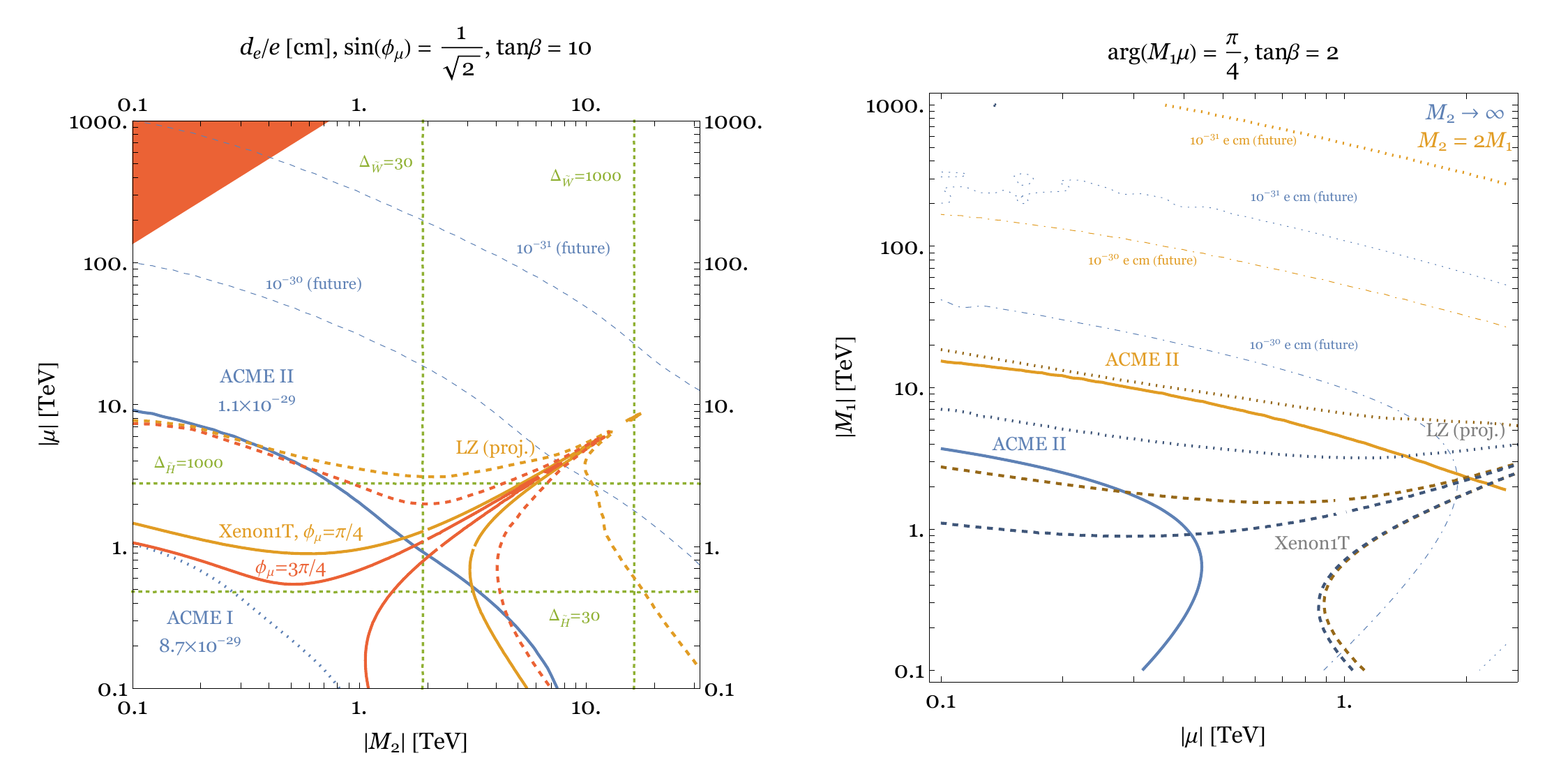}
  \caption{Constraints on electroweakinos from EDMs and from dark matter direct detection in the case of a large CP-violating phase, $\sin(\phi_\mu) = 1/\sqrt{2}$. Left-hand panel: bounds as a function of $|M_2|$ and $|\mu|$, assuming $M_1 = M_2/2$. We have fixed $\tan \beta = 10$ for relatively weak EDM constraints. The orange Xenon1T and LZ curves are for $\phi_\mu = \pi/4$ while the red curves are for $\phi_\mu = 3\pi/4$, where the direct detection constraints are weaker. We see that the EDM constraint is generally stronger except near the diagonal. The green dashed curves are fine-tuning contours and the upper-left triangular region requires tuning away a threshold correction to $M_2$; see \cite[\S6.1]{Nakai:2016atk} for further discussion. The dashed ``future'' curves represent hypothetical future improvements, possibly arising from experiments with polyatomic molecules \cite{Kozyryev:2017cwq}. Right panel: bounds as a function of $|\mu|$ and $|M_1|$ with $\tan \beta = 2$. Here we present two scenarios, one with $M_2 = 2M_1$ and one where winos are decoupled ($M_2 \to \infty$, see \cite{Fox:2014moa}). Decoupling the winos removes the dominant ($\gamma h$) Barr-Zee contributions to the EDM and leaves a much weaker constraint from the $W$ boson EDM. In that case we see that dark matter experiments more strongly constrain the parameter space.}
  \label{fig:ewkino1}
\end{figure}

\begin{figure}[h]
  \centering
  \includegraphics[width=0.4\textwidth]{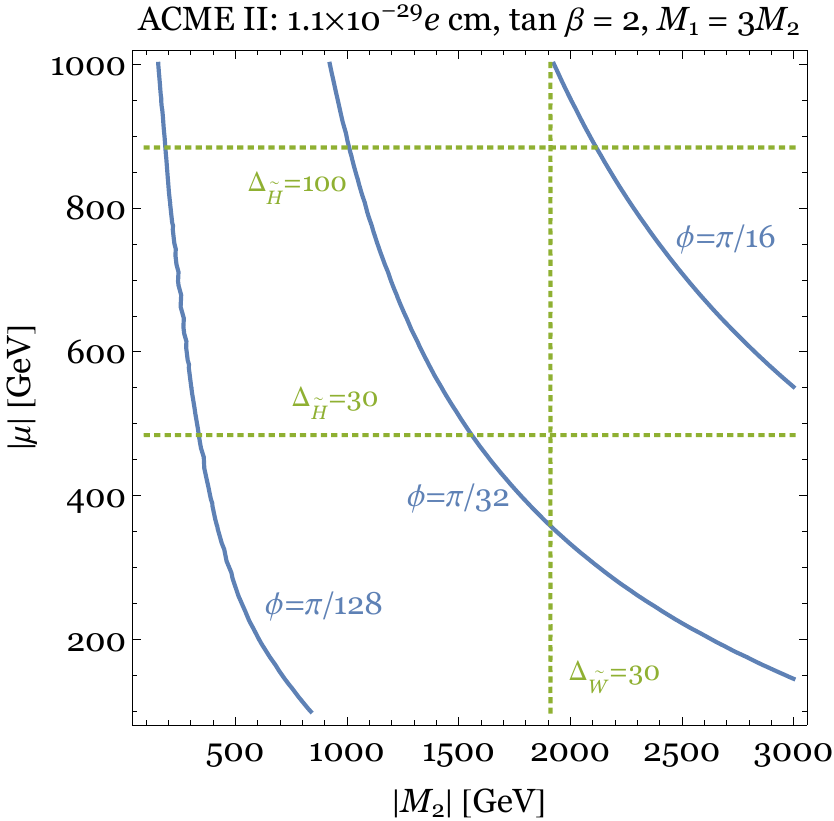}
  \caption{Constraints on electroweakinos from EDMs: the case of small phases. Here we have fixed $M_1 = 3M_2$, as in anomaly mediated SUSY breaking, and $\tan \beta = 2$. We see that the ACME result is compatible with light electroweakinos only for percent-level phases.}
  \label{fig:ewkino2}
\end{figure}

The EDM requires a coupling to the Higgs boson, meaning that it vanishes if we send either the gaugino masses $M_{1,2}$ or the higgsino mass $\mu$ to infinity. As a result, the size of the EDM is highly correlated with a variety of other observables, including the dark matter direct detection cross section if the lightest neutralino is dark matter (see e.g.~\cite{Nagata:2014wma, Krall:2017xij} for further discussion of the EDM/DM interplay). Majorana neutralinos have a dominant spin-independent scattering rate through their coupling to the Higgs boson \cite{Barbieri:1988zs}, which is highly constrained by xenon-based dark matter experiments like Xenon1T \cite{Aprile:2018dbl} and PandaX II \cite{Cui:2017nnn}. Of course, dark matter direct detection experiments can only constrain new physics if the particles in question actually constitute dark matter. In the discussion here we will consider only neutralinos that saturate the observed dark matter relic abundance. (In particular, we do not assume that neutralinos are thermal relics; nonthermal mechanisms for populating dark matter are ubiquitous in SUSY theories.) Neutralinos making up a subdominant fraction of dark matter are more weakly constrained.

In Fig.~\ref{fig:ewkino1} we show comparisons of the electron EDM constraint on electroweakino parameter space with the dark matter direct detection constraint from Xenon1T \cite{Aprile:2018dbl} and a projected future constraint corresponding to the goal of the LZ experiment \cite{Mount:2017qzi}. (For the nucleon matrix elements used in the direct detection calculation we follow \cite{Basirnia:2016szw}, which in turn uses \cite{Belanger:2013oya}.) We see that both experiments are powerful probes of electroweakino masses, reaching into regions of multi-TeV mass. In all curves we have taken the phase appearing in the EDM to be $\phi_\mu \equiv \arg(\mu M_2 b_\mu^*) = \pi/4$ and assumed the phases of $M_1$ and $M_2$ to be equal. In the left-hand panel, we fix $M_1 = M_2/2$ and $\tan \beta = 10$. The right panel of Fig.~\ref{fig:ewkino1} gives a different look at the constraints, focusing on the bino/higgsino sector. The orange curves assume $M_1 = \frac{1}{2} M_2$ as in the left panel, but the light blue curves correspond to the case of a completely decoupled wino ($M_2 \to \infty$), as in the Hypercharge Impure model of Split Dirac SUSY \cite{Fox:2014moa}. This case is of interest for the possibility of nearly pure Dirac higgsino dark matter; we see from the figure that ACME's constraint is relatively weak, though for weak-scale higgsino masses it still probes multi-TeV bino masses. Finally, in Fig.~\ref{fig:ewkino2} we zoom in on the low-mass region of electroweakino parameter space, showing that compatibility with the ACME bound requires small (10\% or lower) phases in the region with a chargino below 1 TeV.

Although we have focused in this section on charginos and neutralinos in the SUSY context, so that the Yukawa couplings are pinned to the size of Standard Model gauge interactions, much of the discussion would carry over to a more general scenario with new fermions with electroweak quantum numbers and Yukawa couplings to the Higgs boson. These are often discussed in the dark matter context as singlet-doublet and doublet-triplet models \cite{Mahbubani:2005pt,  Cohen:2011ec, Dedes:2014hga}. The interplay between EDM constraints and other probes of such fermions as dark matter has been discussed in \cite{Fan:2013qn}. If new fermions with large Yukawa couplings are added to the Standard Model without additional bosons (such as their supersymmetric partners), radiative corrections can destabilize the Higgs potential and lead to rapid vacuum decay \cite{Joglekar:2012vc, ArkaniHamed:2012kq, Blum:2015rpa}. As a result such particles are often discussed in the context of supersymmetry \cite{Joglekar:2013zya}, and are interesting for explaining the 125 GeV Higgs mass \cite{Basirnia:2016szw}. In the presence of CP violation, these models also provide an appealing fermionic scenario for electroweak baryogenesis \cite{Egana-Ugrinovic:2017jib, Carena:2004ha}.

%%%%%%%%%%%%%%%%%%%%%%%%%%%%%%%%%%%%%%%%%%%%%%
\section{The EDM constraint on natural SUSY}
\label{sec:twoloopnatural}

\begin{figure}[!t]
 \begin{minipage}{0.45\hsize}
  \begin{center}
   \includegraphics[clip, width=7.5cm]{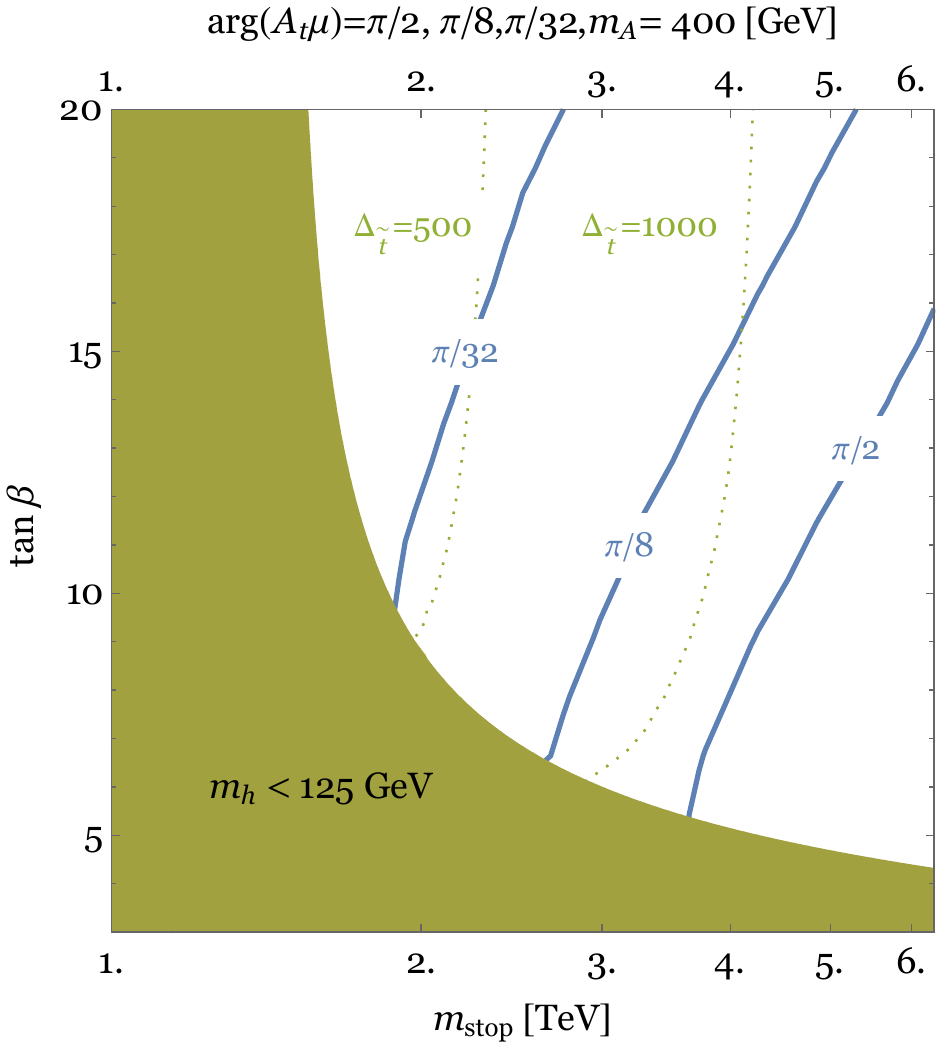}
  \end{center}
 \end{minipage}
 \hspace{1cm}
 \begin{minipage}{0.45\hsize}
  \begin{center}
   \includegraphics[clip, width=7.6cm]{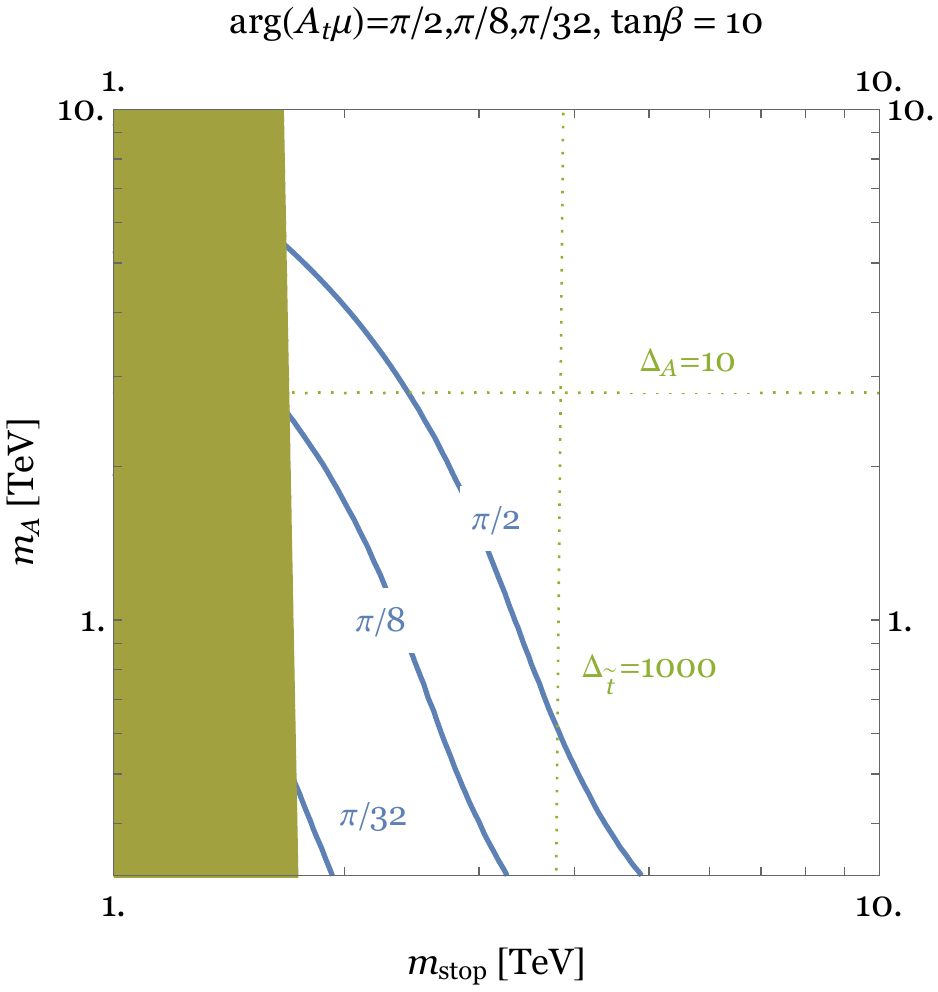}
  \end{center}
 \end{minipage}
  \caption{The implication of the EDM bound in the ACME II experiment on the stop parameter space in the MSSM
  where the 125 GeV Higgs mass is realized by stop loops with a large $A$-term.
  The horizontal axis is the common stop mass $m_{\rm stop} = {m}_{\widetilde{Q}_3} ={m}_{ \widetilde{u}_3}$.
  The vertical axes show $\tan \beta$ and $m_A$ in the left and right panels respectively.
  We fix $m_A = 400 \, \rm GeV$ in the left panel and $\tan \beta = 10$ in the right panel.
  The phase is taken to be $\arg (A_t \mu) = \pi/2, \pi/8, \pi/32$.
  The parameter $|\mu|$ is $350 \, \rm GeV$. 
  The green region is excluded by the small Higgs mass with any values of the $A$-term.
The blue curves denote the ACME II constraint.
The green dotted curve describes the degree of fine-tuning defined in \eqref{tuningmeasure}.
  }
  \label{fig:mssmEDM}
\end{figure}

\begin{figure}[!t]
 \begin{minipage}{0.45\hsize}
  \begin{center}
   \includegraphics[clip, width=7.5cm]{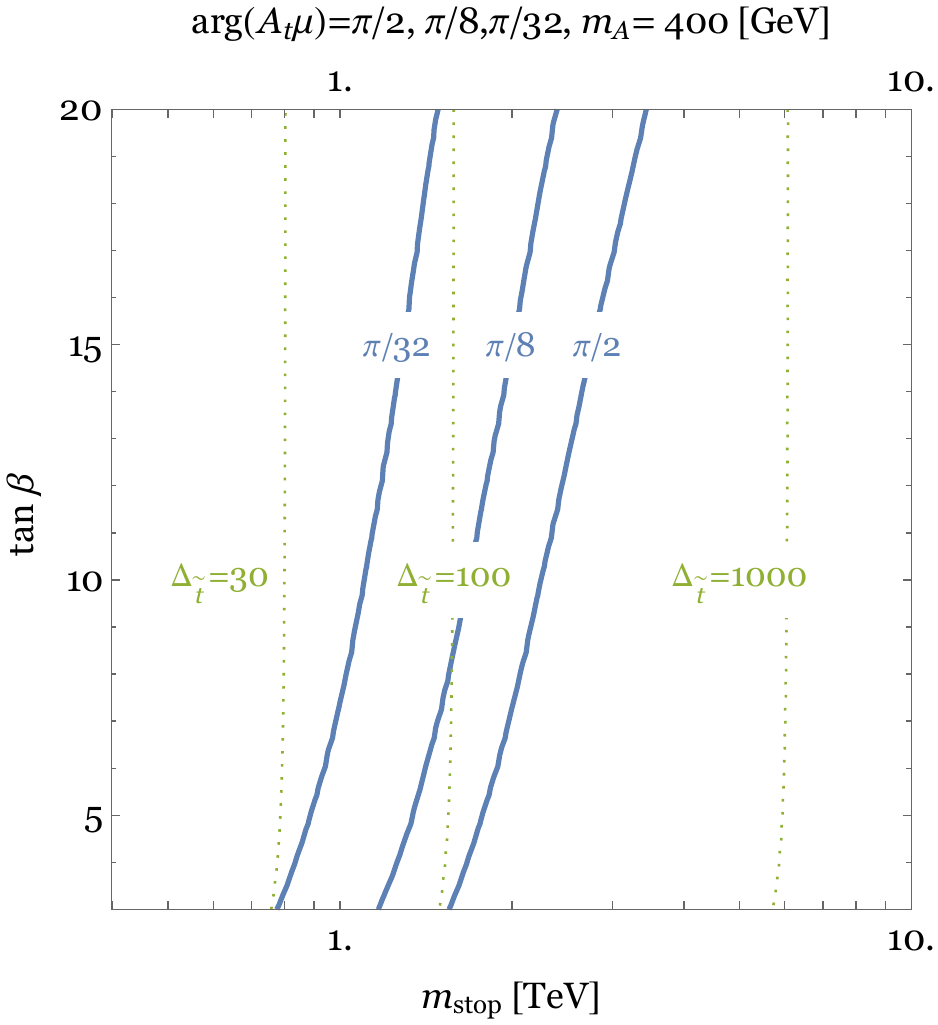}
  \end{center}
 \end{minipage}
 \hspace{1cm}
 \begin{minipage}{0.45\hsize}
  \begin{center}
   \includegraphics[clip, width=7.6cm]{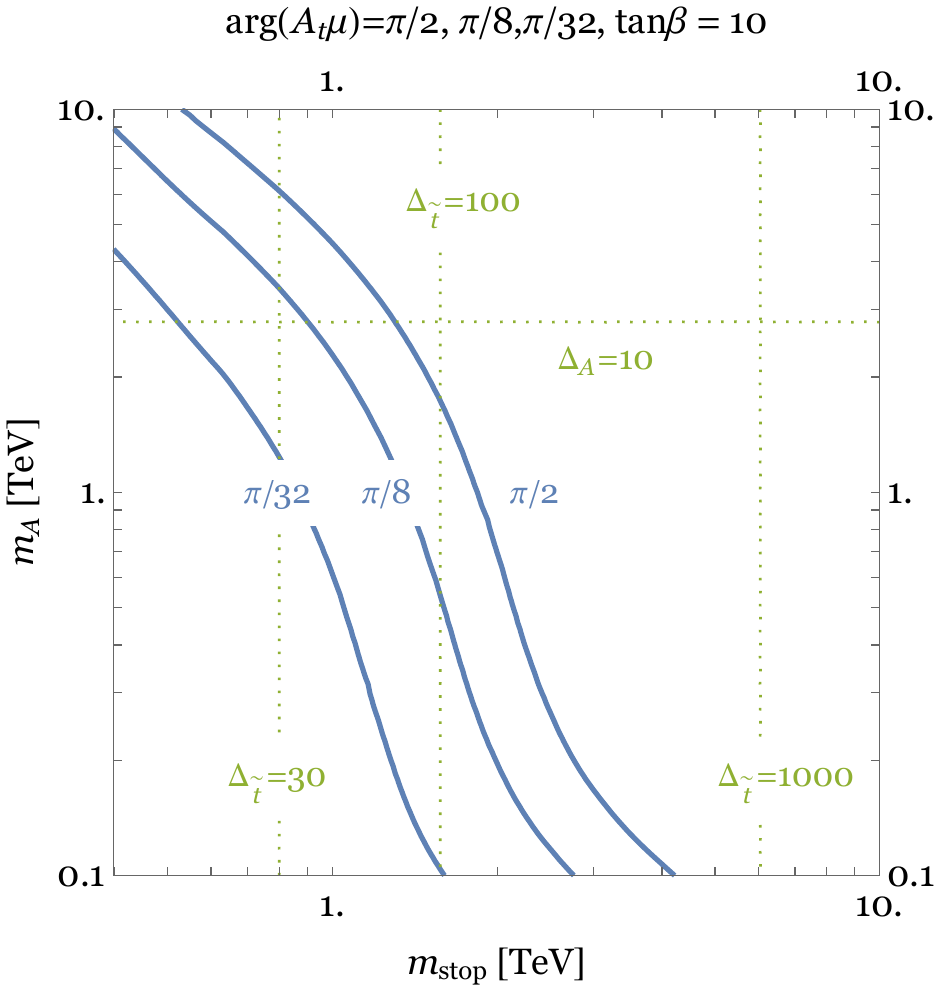}
  \end{center}
 \end{minipage}
  \caption{The implication of the EDM bound in the ACME II experiment on the stop parameter space assuming some other interactions
  to explain the correct Higgs mass.
  The $A$-term is still radiatively generated from the gluino mass.
  The parameters are taken to be the same values as those of figure~\ref{fig:mssmEDM}.
  The blue curves denote the ACME II constraint with phases $\arg (A_t \mu) = \pi/2, \pi/8, \pi/32$.}
  \label{fig:EDM_Dterm}
\end{figure}

In this section, we explore implications of the EDM constraint from the ACME II experiment on the framework of natural SUSY
where the only particles that play a key role in naturalness of the electroweak breaking are relatively light.
That is, only higgsinos, stops, the left-handed sbottom and gauginos are light and the other superpartners such as the first and second generations of squarks and sleptons
can be very heavy.
While the squarks and sleptons contribute to sizable FCNCs and 1-loop EDMs,
these contributions are suppressed in natural SUSY.
In this framework, the two-loop Barr-Zee-type diagram, with an inner stop loop connected to the electron with a photon and a pseudoscalar Higgs, gives the dominant contribution to the electron EDM.
This scenario gives a minimum size of the EDM in TeV-scale SUSY with CP-violation whether or not 1-loop contributions from the squarks or sleptons exist or not.

The MSSM predicts a small Higgs mass because the tree-level Higgs quartic coupling is related to the electroweak gauge couplings.
Then, a sizable radiative correction is needed to explain the 125 GeV Higgs mass.
In general, stop masses must be around 10 TeV, which leads to a serious fine-tuning.
However, if we consider a near-maximal stop mixing,
the correct Higgs mass can be realized with light stops by a large $A$-term.
The two-loop EDM induced by stops is proportional to the $A$-term and we obtain a detectably large EDM with a nonzero phase
of the $A$-term, $\arg (A_t \mu)$.
Another direction to realize light stops and reduce the tuning is to extend the Higgs sector of the MSSM and provide a new interaction to lift up the Higgs mass.
In this case, we do not need a large $A$-term (and heavy stops), but
the $A$-term is still radiatively generated from the gluino mass, which can lead to a nonzero EDM.

Let us now investigate implications on the natural SUSY parameter space from the ACME II experiment.
We consider two scenarios to raise the Higgs mass described above.
Figure~\ref{fig:mssmEDM} shows the stop parameter space in the MSSM
where the 125 GeV Higgs mass is realized by stop loops with a large $A$-term.
We use the SusyHD code~\cite{Vega:2015fna} for the Higgs mass calculation.
We assume the same left and right-handed stop masses $m_{\rm stop} = {m}_{\widetilde{Q}_3} = {m}_{\widetilde{u}_3}$
for simplicity.
The parameter $|\mu|$ is taken to be $350 \, \rm GeV$. 
The green region is excluded by the small Higgs mass with any values of the $A$-term.
The blue curves denote the ACME II constraint.
We estimate the degree of tuning by using the following measures (for a more detailed discussion, see \cite{Katz:2014mba,Nakai:2016atk}),
\begin{equation}
\begin{split}
\Delta_A \equiv \frac{2 m_A^2}{m_h^2 \tan^2 \beta} , \qquad \quad \Delta_{\tilde{t}} \equiv \left| \frac{2 \delta m_{H_u}^2}{m_h^2} \right| ,
\label{tuningmeasure}
\end{split}
\end{equation}
where $m_A$ is the pseudoscalar Higgs mass and $\delta m_{H_u}^2$ denotes the stop radiative correction to the up-type Higgs soft mass squared.
The degree of fine-tuning from the stop radiative correction is worse than one percent in this scenario.

Figure~\ref{fig:EDM_Dterm} shows
the EDM constraint on the stop parameter space assuming some other interactions
to explain the correct Higgs mass.
As described above, the $A$-term is still radiatively generated from the gluino mass.
The parameters are taken to be the same values as those of figure~\ref{fig:mssmEDM}.
The blue curves denote the ACME II constraint.
In the viable parameter region, at least one percent tuning is needed.

\section{QULE-induced EDMs}
\label{sec:QULE}

\subsection{The RGE of the electron EDM}
\label{subsec:RGE_qule}

The electron EDM could be induced from various other dimension-six operators in the Standard Model through renormalization group equations. The relevant RGE has been given in \cite{Alonso:2013hga} (which together with \cite{Jenkins:2013zja,Jenkins:2013wua} constructs the RGEs for all of the SM dimension-six operators \cite{Grzadkowski:2010es}).
In a standard chosen basis of dimension six operators, the subset which are of interest for the measurement of CP violation through leptons is:
\begin{align}
{\cal L}_{\rm dim\, 6} &\supset C_{h {\widetilde W} B} h^\dagger \sigma^i h {\widetilde W}^i_{\mu \nu} B^{\mu \nu} + C_{h{\widetilde W}} h^\dagger h {\widetilde W}^i_{\mu \nu} W^{i \mu \nu} + C_{h{\widetilde B}} h^\dagger h \widetilde{B}_{\mu \nu} B^{\mu \nu} \nonumber \\[1ex]
& + \left[C_{\substack{eW}} ({\bar \ell} \sigma^{\mu \nu} e ) \sigma^i h W^i_{\mu \nu} + C_{\substack{eB}} ({\bar \ell} \sigma^{\mu \nu} e) h B_{\mu \nu} + {\rm h.c.}\right] \nonumber \\[1ex]
& + \left[C_{ledq; f} (\bar{\ell} e) \cdot (\bar{d}_f q_f) + C^{(1)}_{\substack{lequ; f}} ({\bar \ell} e) \cdot ({\bar q}_f u_f) + C^{(3)}_{\substack{lequ; f}} ({\bar \ell} \sigma_{\mu \nu} e) \cdot ({\bar q}_f \sigma^{\mu \nu} u_f) + {\rm h.c.} \right].   \label{eq:operatorlist}
\end{align}
We see several four-fermion operators in the last line. Only one of them, the operator $(q_f \sigmabar^{\mu \nu}{\bar u}_f) \cdot (\ell {\sigmabar}_{\mu \nu} {\bar e})$, which we will refer to as the ``QULE operator,'' feeds into the electron EDM through the 1-loop RGE. On the other hand, the first two types of operators, whose coefficients are $C_{ledq; f}$ and $C^{(1)}_{\substack{lequ; f}}$,  contribute to the CP-odd electron-nucleon coupling $C_S$ that we discussed in \S\ref{subsec:CScontrib}.
We can easily extract the coefficients $C_{qe}$ in \eqref{fourfermiint} from these operators.

The RGE in \cite{Alonso:2013hga} is given for a quantity $\widetilde{\cal C}_{e \gamma}$ which is, in our conventions, $-\sqrt{2} d_e/(e v)$. The one-loop RGE shows that the EDM can be generated from:
\begin{itemize}
\item The electron EDM $\widetilde{\cal C}_{e \gamma}$ or the related $\widetilde{\cal C}_{e Z}$ which replaces the photon coupling with a coupling to the $Z$-boson. When these contributions appear, we just have a one-loop renormalization of a pre-existing EDM.
\item The Wilson coefficients $C_{h {\widetilde W} B}$, $C_{h \widetilde{W}}$, and $C_{h \widetilde{B}}$. If these arise at one loop, then the two-loop EDM they generate is of the well-studied Barr-Zee type.
\item The Wilson coefficient ${\rm Im}\, C^{(3)}_{\substack{lequ; f}}$, i.e.~the ``QULE operator.''
\end{itemize}
It is the last of these contributions, which has received relatively little attention in the literature, that we turn our attention to now. The EDM arises diagramatically by closing up the quark loop with a Higgs insertion and attaching a photon line. (We make the simplifying assumption that quarks appear in a flavor-diagonal manner.) The leading-log estimate of the EDM induced by the quark flavor $f$ is given by
\be
d_e = -e \frac{m_f}{\pi^2} \log \frac{M}{m_f}\, {\rm Im}\, C^{(3)}_{lequ; f}, \label{eq:EDMfromQULE}
\ee
with $m_f$ the mass of the quark flavor appearing in the loop and $M$ the scale at which the QULE operator is generated. 

Some models of new physics will generate four-fermion operators that are not expressed in the chosen basis, so we must make use of identities to reduce to this basis to determine if the QULE Wilson coefficient ${\rm Im}\, C^{(3)}_{lequ; f}$ is nonzero. By Fierz rearrangement, we see that the four-fermi operators $({\bar \ell}u) \cdot ({\bar q}e)$ and $({\bar \ell}\cdot {\bar q})(ue)$ generate an EDM although the operator $({\bar \ell}e) \cdot ({\bar q}u)$ does not (though it does contribute to the CP-odd electron-nucleon term $C_S$ discussed in \S\ref{subsec:CScontrib}). Specifically, we have
\begin{align}
(q {\bar e}) \cdot (\ell {\bar u}) &= -\frac{1}{2} \left[ (q \sigmabar^{\mu \nu} {\bar u}) \cdot (\ell \sigmabar_{\mu \nu} {\bar e}) + (q {\bar u}) \cdot (\ell {\bar e})\right], \nonumber \\
(q \cdot \ell) ({\bar u}{\bar e}) &= +\frac{1}{2} \left[ (q \sigmabar^{\mu \nu} {\bar u}) \cdot (\ell \sigmabar_{\mu \nu} {\bar e}) - (q {\bar u}) \cdot (\ell {\bar e})\right]. \label{eq:QULEfierz}
\end{align}
Thus, models that generate $(q {\bar e}) \cdot (\ell {\bar u})$ or $(q \cdot \ell) ({\bar u}{\bar e})$ will, when rewritten in the operator basis of \cite{Alonso:2013hga}, have a QULE contribution. In this section we will survey models of QULE-induced EDMs. The first case is a one-loop EDM arising from a tree-level QULE operator. The second case is a two-loop EDM arising when a QULE operator is generated through a one-loop box diagram.

The two 4-fermi operators in \eqref{eq:QULEfierz} generate both $C^{(3)}_{lequ; f}$ (and hence, at one higher loop order, an EDM) as well as $C^{(1)}_{lequ; f}$, which contributes to $C_S$ as discussed in \S\ref{subsec:CScontrib}. Since ACME constrains the linear combination $d_{\rm ThO}$ in \eqref{eq:dThO}, we can ask whether the constraint on these operators is dominantly from $d_e$ or from $C_S$. The ratio of the two contributions is
\begin{align}
\left|\frac{d_{\rm ThO; EDM}}{d_{{\rm ThO;} C_S}}\right| &\approx \frac{m_q \log(M/m_q)}{\pi^2 \times 1.6 \times 10^{-15}~{\rm GeV}^2~{\rm cm}~\langle N | {\bar q} q | N \rangle} \nonumber \\
&\approx \begin{cases} 
                       6 \times 10^{-3}, & q = u, \\
                       2 \times 10^{2}, & q = c, \\
                       2 \times 10^{6}, & q = t.
                \end{cases}
\label{eq:de_to_Cs}
\end{align}
In the numerical estimate we have plugged in $M = 10~{\rm TeV}$ for the mass scale running in the loop, though the result depends only logarithmically on this choice. The upshot is that if new physics couples dominantly to a heavier up-type quark, the constraint is primarily on the electron EDM, while if new physics couples dominantly to the up quark, the constraint is primarily on the CP-odd electron-nucleon interaction $C_S$. This is consistent with results in \cite{Fuyuto:2018scm}. Of course, new physics might couple to all of the quarks, in which case the flavor structure of the interaction will determine which quark gives the largest contribution.

\subsection{One loop EDM from tree level QULE}\label{subsec:treeQULE}

A nonzero Wilson coefficient $C^{(3)}_{lequ; f}$ (after reducing to the appropriate basis) can be generated at tree level by integrating out a scalar with leptoquark-type couplings. In such a model, there is a one-loop electron EDM \cite{Arnold:2013cva, Dorsner:2016wpm, Fuyuto:2018scm,Dekens:2018bci}. There are two possible charge assignments for the scalar that will generate a QULE operator that induces an EDM:
\begin{align}
\phi \in ({\bf 3}, {\bf 1})_{-1/3}, \quad & {\cal L} \supset \left(y_{1f} \phi^\dagger q_f \cdot \ell + y_{2f} \phi {\bar u}_f {\bar e} + {\rm h.c.}\right) - m_\phi^2 \phi^\dagger \phi, \\
{\rm generates:} \quad & C^{(3)*}_{lequ; f} = - C^{(1)*}_{lequ; f} = \frac{y_{1f}y_{2f}}{2 m_\phi^2}, \\
\phi \in ({\bf 3}, {\bf 2})_{+7/6}, \quad & {\cal L} \supset \left(y_{1f} \phi^\dagger q_f {\bar e} + y_{2f} \phi \cdot \ell {\bar u}_f + {\rm h.c.}\right) - m_\phi^2 \phi^\dagger \phi, \\
{\rm generates:} \quad & C^{(3)*}_{lequ; f} = C^{(1)*}_{lequ; f} = -\frac{y_{1f}y_{2f}}{2 m_\phi^2}.
\end{align}
The case $({\bf 3}, {\bf 1})_{-1/3}$ (the quantum numbers of a down-type squark) allows for diquark-like couplings $\phi q_f \cdot q_g$ and $\phi^\dagger {\bar u}_f {\bar d}_g$, which together with the couplings above violate baryon number and can lead to proton decay. Hence, this case requires some mechanism (or an extreme accident) to suppress these dangerous couplings. However, the case of $({\bf 3}, {\bf 2})_{+7/6}$ does not share this problem. Both models generate a contribution to $C_S$ and to $d_e$ (our result appears to differ from \cite{Fuyuto:2018scm}, which claims that only the case $({\bf 3}, {\bf 2})_{+7/6}$ generates both operators).
In the case where the scalar $\phi$ couples to the top quark, the constraint that loop corrections do not generate a large correction to the electron Yukawa coupling is $|y_{1t} y_{2t}| \lesssim \mathcal{O}(10^{-6})$. A variety of assumptions about the flavor structure of the model are possible: in some models $\phi$ may couple most strongly to the third generation, while in others it may couple to all generations of quarks. In any case, the electron mass naturalness constraint prevents the couplings from being too large. 
In the case where the scalar $\phi$ couples to the up quark, the CP-odd electron-nucleon coupling $C_S$ leads to the strongest constraint on the model. 
The scalar particle in this model must be very heavy and far out of reach for collider searches unless it has small couplings.

\subsection{Two loop EDM from one loop QULE}

\begin{figure}[h]
  \centering
  \includegraphics[width=1.0\textwidth]{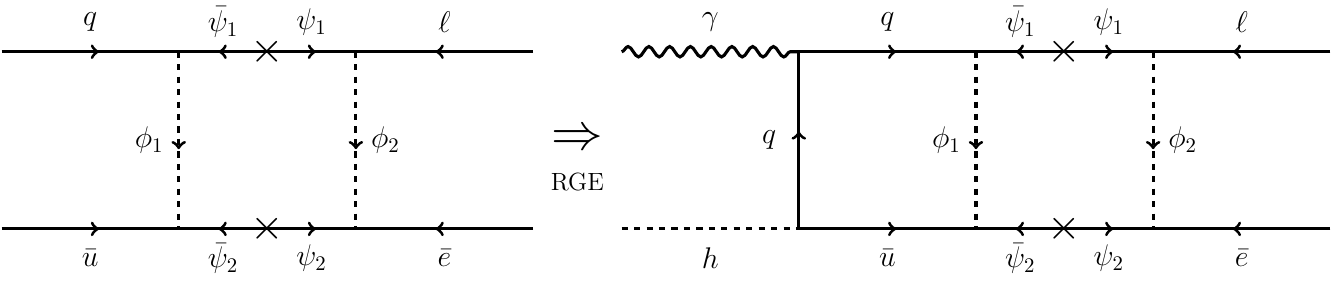}
  \caption{Feynman diagrams for an EDM arising at two loops from a one-loop QULE operator. Similar diagrams exist that generate the operator $(q {\bar e}) \cdot ({\bar u} \ell)$ instead of $(q \cdot \ell)({\bar u}{\bar e})$.}
  \label{fig:feynmanQULE}
\end{figure}

We could also consider theories in which a QULE-type operator is generated, not at tree level, but at one loop. In this case the corresponding EDM will arise from a 2-loop diagram, as depicted in Fig.~\ref{fig:feynmanQULE}. These diagrams are topologically distinct from Barr-Zee diagrams in that they do not contain a closed internal fermion loop; rather, a single fermion line runs continuously through the diagram. (The diagrams are also different from previously discussed rainbow diagrams which have Standard Model fermions on some internal lines \cite{Yamanaka:2012ia}.) As shown in the figure, such diagrams can arise if we introduce new vectorlike fermions $\psi_{1,2}, {\bar \psi}_{1,2}$ and complex scalars $\phi_1, \phi_2$ with appropriate Yukawa couplings. A variety of choices of quantum numbers for the particles are possible. For simplicity, we make the simplifying assumption that a single $SU(3)$ color index and $SU(2)$ weak isospin index run continuously through the diagram, e.g.~in Fig.~\ref{fig:feynmanQULE} we might take $\phi_1$ to be a color singlet and $\psi_1, \phi_2, {\bar \psi}_2$ to be color triplets. With such an assumption, it is a straightforward (but lengthy) task to enumerate all of the possibilities. We provide these results in Appendix \ref{app:QULE}.

In some cases, models with these contributions also allow couplings that generate contributions of more well-studied types. For example, from Fig.~\ref{fig:feynmanQULE} we see that the quantum numbers allow for a vertex connecting $\psi_1$ and $\psi_2$ to a Higgs boson, which would allow a 1-loop electron EDM. If this coupling to the Higgs is present, then the 1-loop EDM could dominate over the QULE contribution we study. We could {\em assume} the Higgs coupling to be small, so that the QULE contribution is dominant, but this assumption is not obviously well-motivated. On the other hand, if we exchanged the position of $\bar u$ and $\bar e$ in the lower line of Fig.~\ref{fig:feynmanQULE}, then in general the quantum numbers of the new particles would not allow any direct couplings to the Higgs boson. In that case both 1-loop EDMs and 2-loop Barr-Zee EDMs would be absent, and the QULE contribution would be dominant.

Rather than directly computing a 2-loop diagram as in the right panel of Fig.~\ref{fig:feynmanQULE}, we consider the leading-log approximation given by feeding the box diagram in the left panel into the RGE estimate \eqref{eq:EDMfromQULE}. The coefficient of the four-fermion operator generated by this box diagram is given by
\begin{equation}
\begin{split}
C_{\rm g} \cdot y_1 y_2 y_3 y_4 \int \frac{d^4 p}{(2\pi)^4} \frac{1}{p^2 - m_{\phi_1}^2} \frac{m_{\psi_1}}{p^2 - m_{\psi_1}^2} \frac{1}{p^2 - m_{\phi_2}^2} \frac{m_{\psi_2}}{p^2 - m_{\psi_2}^2},
\end{split}
\end{equation}
with $C_{\rm g}$ a group-theory factor depending on the $SU(3)_C$ and $SU(2)_L$ representations of the particles running in the loop. With our simplifying assumption about representations above, the group theory factor is 1.
Evaluating the loop integral, we obtain
\begin{equation}
\begin{split}
&\frac{C_{\rm g} y_1 y_2 y_3 y_4 m_{\psi_1} m_{\psi_2}}{16\pi^{2}(m_{\psi_1}^{2}-m_{\psi_2}^{2})(m_{\psi_1}^{2}-m_{\phi_2}^{2})(m_{\psi_2}^{2}-m_{\phi_2}^{2})(m_{\psi_1}^{2}-m_{\phi_1}^{2})(m_{\psi_2}^{2}-m_{\phi_1}^{2})(m_{\phi_1}^{2}-m_{\phi_2}^{2})}\times\\
&\left\{m_{\phi_1}^2m_{\psi_1}^2(m_{\psi_2}^4-m_{\phi_2}^4)\log\frac{m_{\phi_1}^2}{m_{\psi_1}^2}+m_{\phi_1}^2m_{\phi_2}^2(m_{\psi_1}^4-m_{\psi_2}^4)\log\frac{m_{\phi_1}^2}{m_{\phi_2}^2}+m_{\phi_1}^2m_{\psi_2}^2(m_{\phi_2}^4-m_{\psi_1}^4)\log\frac{m_{\phi_1}^2}{m_{\psi_2}^2}\right.\\
&\left.+m_{\psi_1}^2m_{\psi_2}^2(m_{\phi_2}^4-m_{\phi_1}^4)\log\frac{m_{\psi_2}^2}{m_{\psi_1}^2}+m_{\phi_2}^2m_{\psi_1}^2(m_{\psi_2}^4-m_{\phi_1}^4)\log\frac{m_{\psi_1}^2}{m_{\phi_2}^2}+m_{\phi_2}^2m_{\psi_2}^2(m_{\phi_1}^4-m_{\psi_1}^4)\log\frac{m_{\psi_2}^2}{m_{\phi_2}^2}\right\}.
\end{split}
\end{equation}

\begin{figure}[h]
  \centering
  \includegraphics[width=0.6\textwidth]{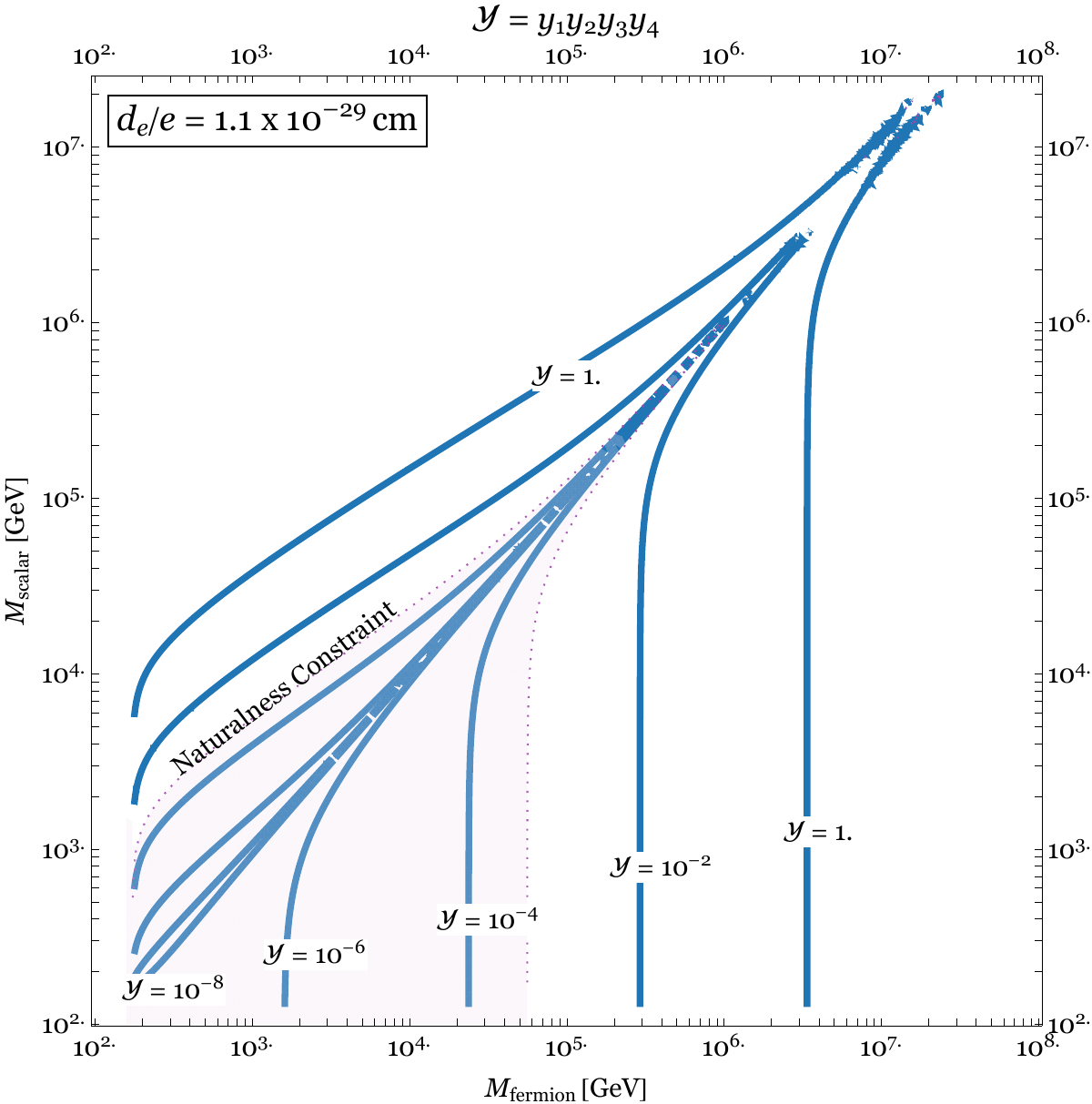}
  \caption{The constraint of the electron EDM arising at two loops from a one-loop QULE operator. Contours show the largest allowed imaginary part of the product of Yukawa couplings appearing in the box diagram, as a function of the masses of the fermions and scalars in the loop. The shaded region shows the case where the constrained value of $\mathcal Y$ is small enough that it does not generate an unnaturally large correction to the electron Yukawa coupling. We see that new physics as heavy as a few hundred TeV can be constrained, consistent with the estimate in the radiative scenario in \S\ref{sec:bigpicture}.}
  \label{fig:loopQULE}
\end{figure}

While we reserve the detailed discussion of possible quantum numbers for the appendix \ref{app:QULE}, let us highlight some example scenarios here.
\begin{itemize}
\item {\bf SUSY:} in Fig.~\ref{fig:feynmanQULE}, we could take $\phi_1 = {\widetilde u}$; ${\bar \psi}_1 = {\widetilde H}_u$; $\psi_1 = {\widetilde H}_d$; $\phi_2 = {\widetilde e}$; and ${\bar \psi}_2 = \psi_2 = \widetilde{B}^0$. Then the diagram gives rise to a 2-loop SUSY EDM, distinct from the Barr-Zee type diagram, and scaling approximately as $\sim g_1^2 y_e y_t^2 \mu M_1/\left[(16\pi^2)^2 m_{\widetilde t}^2 m_{\widetilde e}^2 \right] \log(m_\phi/m_t)$. Such a contribution is generally expected to be smaller than the one-loop diagrams considered in \S\ref{sec:oneloopSUSY}.
\item {\bf New physics parities:} some possibilities resemble SUSY in having an analogue of $R$-parity, with all of the new physics particles in the loop charged under a parity symmetry so that they can cascade decay to a neutral parity-odd particle (which could serve as a dark matter candidate). For example, consider this scenario (the $Y = \frac{1}{6}$ row of table \ref{sec:A11tab})
     \begin{equation}
     \begin{tikzpicture}[line width=1.5 pt]
	\draw[fermion] (-4.5,0)--(-1.5,0);
	\node at (-3,-0.5) {$\bar{e}\ (1, 1)_{1}$};
	\draw[fermion] (4.5,0)--(1.5,0);
	\node at (3,-0.5) {$\bar{u}\ (\bar{3}, 1)_{-\frac{2}{3}}$};
	\draw[fermion] (-1.5,0)--(1.5,0);
	\node at (0, -0.5) {$(3, 2)_{\frac{7}{6}}$};

	\draw[scalar] (-1.5, 3) -- (-1.5, 0);
	\node at (-2.5, 1.5) {$(3, 2)_{\frac{1}{6}}$};
	\draw[scalar] (1.5, 3)  -- (1.5, 0);
	\node at (2.5, 1.5) {$(1, 2)_{-\frac{1}{2}}$};

	\draw[fermion] (-4.5,3)--(-1.5,3);
	\node at (-3,3.5) {$q\ (3, 2)_{\frac{1}{6}}$};
	\draw[fermion] (4.5,3)--(1.5,3);
	\node at (3,3.5) {$\ell\ (1, 2)_-\frac{1}{2}$};
	\draw[fermion] (-1.5,3)--(1.5,3);
	\node at (0, 3.5) {$(1, 1)_{0}$};
	\end{tikzpicture}
     \end{equation}
This case includes an exotic vectorlike doublet quark $X$ of hypercharge $7/6$, corresponding to electric charges $5/3$ and $2/3$. This particle can decay as, for example, $X_{+5/3} \to t e^+ \psi^0$ with $\psi^0$ a neutral stable fermion. Thus we see that the collider signals for some scenarios resemble SUSY in having missing momentum, but resemble leptoquarks in having decay chains with both quarks and leptons. The final state for $X$ pair production is $t{\bar t}e^+ e^- + p^{\rm miss}_T$. Because two different decay chains with the same final state are open (moving clockwise or counterclockwise from the $(3,2)_{7/6}$ particle to the $(1,1)_0$ particle in the diagram), reconstructing masses of the intermediate scalars could be an interesting challenge.

Notice that this is an example in which the two-loop QULE contribution is the leading possible EDM, because no couplings permitting a one-loop EDM or Barr-Zee diagram exist.

\item {\bf Leptoquarks:} 
In some cases, the quantum numbers permit one of the particles running in the loop to decay to one Standard Model quark and one lepton. For example, consider the $Y = 0$ row of \ref{sec:A41tab}:
     \begin{equation}
	\begin{tikzpicture}[line width=1.5 pt]
	\draw[fermion] (-4.5,0)--(-1.5,0);
	\node at (-3,-0.5) {$\bar{u}\ (\bar{3}, 1)_{-\frac{2}{3}}$};
	\draw[fermion] (4.5,0)--(1.5,0);
	\node at (3,-0.5) {$\ell\ (1, 2)_{-\frac{1}{2}}$};
	\draw[fermion] (-1.5,0)--(1.5,0);
	\node at (0, -0.5) {$(\bar{3}, 1)_{-\frac{2}{3}}$};

	\draw[scalar] (-1.5, 3) -- (-1.5, 0);
	\node at (-2.5, 1.5) {$(1, 1)_{0}$};
	\draw[scalar] (1.5, 3)  -- (1.5, 0);
	\node at (2.5, 1.5) {$(3, 2)_{\frac{7}{6}}$};

	\draw[fermion] (-4.5,3)--(-1.5,3);
	\node at (-3,3.5) {$q\ (3, 2)_{\frac{1}{6}}$};
	\draw[fermion] (4.5,3)--(1.5,3);
	\node at (3,3.5) {$\bar{e}\ (1, 1)_1$};
	\draw[fermion] (-1.5,3)--(1.5,3);
	\node at (0, 3.5) {$(3, 2)_{\frac{1}{6}}$};
	\end{tikzpicture}
     \end{equation}
In this case, the scalar $\phi_2$ with quantum numbers $({\bf 3}, {\bf 2})_{7/6}$ can decay to one quark and one lepton if appropriate Yukawa couplings exist to $q{\bar e}$ or to $(\ell {\bar u})^\dagger$. The other particles could cascade down to it, for instance, 
   \begin{equation}
   \phi_1 \to u \psi_2, \quad \psi_2 \to {\bar \ell} \phi_2^*, \quad \phi_2^* \to \ell {\bar u},
   \end{equation}
so that the final states could involve several quarks and leptons. An alternative phenomenological scenario for this choice of quantum numbers is that $\psi_1$ or $\psi_2$, a vectorlike quark, is the lightest of the new particles; these could then decay through a Yukawa coupling with a SM quark and the Higgs boson.

This is an example of a case where couplings leading to a 1-loop EDM are allowed to exist, and so the 2-loop contribution considered here may be subdominant.

Scalar leptoquarks are of phenomenological interest for many reasons. For example, a leptoquark with the quantum numbers of a right-handed down squark has been suggested as an explanation for the $B \to D^{(*)}\tau \bar{\nu}$ anomaly \cite{Freytsis:2015qca} and could even fit other flavor anomalies through loop effects \cite{Bauer:2015knc}. 
\end{itemize}

In Fig.~\ref{fig:loopQULE} we plot the experimental constraints on these models. The product of Yukawa couplings $\mathcal{Y} \equiv y_1 y_2 y_3 y_3$ in the one-loop QULE are constrained such that it does not generate an unnaturally large correction to the electron Yukawa coupling. The maximum value of $\mathcal Y$ allowed by naturalness is weakly dependent on the masses of the scalars and fermions in the loop, and coincides roughly with $\mathcal{Y} \sim 10^{-4}$. Contours of fixed $\mathcal{Y} \lesssim 10^{-4}$ lay neatly in the shaded region and are allowed. We see that new physics up to several hundred TeV are consistent with this constraint.

\section{Conclusions}
\label{sec:conclusions}

We have studied implications of the new ACME constraint on a variety of theories of new physics with CP violation. The general argument based on effective field theory has revealed the range of mass scales probed by the EDM constraint. In scenarios with two-loop EDMs where the electron Yukawa coupling appears explicitly in the new physics couplings, including many SUSY scenarios, the new ACME constraint probes masses of a few TeV. Other scenarios, where loop effects generate both the EDM and the electron Yukawa coupling, potentially probe scales of hundreds of TeV.
We have also discussed the case where the dominant effect on ThO comes from the CP-odd electron-nucleon coupling.
Then, we have interpreted the bound in the context of different scenarios for SUSY. For 1-loop SUSY, the constraint probes sleptons above 10 TeV. Assuming the universal mass for squarks and sleptons, the mass bound start to hit the excluded region giving a too-large Higgs mass. For 2-loop SUSY, multi-TeV charginos in split SUSY or stops in natural SUSY are constrained from the new EDM result, which is consistent with
the general argument.

Although there has been extensive study of scenarios where an electron EDM is induced at two loops by new electroweak physics coupling to the Higgs boson, an equally viable possibility is that the electron EDM arises at two loops from physics that is decoupled from the Higgs boson. Such physics, instead, would couple to the charm or (perhaps more plausibly) the top quark. This possibility is realized through the QULE operator which generates the EDM through RG evolution. We have classified scenarios in which the QULE operator is generated at one loop through a box diagram, which include SUSY and leptoquark models. The electron EDM bound gives the leading constraint in most viable models. If a nonzero electron EDM is measured in the future, it will be of paramount importance for colliders to search for the particles responsible for the effect. We have seen that a variety of models with distinctive phenomenology could be the source of the EDM through the QULE operator. In particular, searches for scalar leptoquarks or heavy vectorlike fermions that decay to Standard Model fermions could play a role in pinning down the origin of the EDM if it is measured to be nonzero.

The rest of this section is devoted to a brief discussion on the implications of future improvements in EDM searches. We will discuss first the possibility that null results persist, and then the possibility that a nonzero EDM is measured. We argue that both cases indicate a variety of interesting directions in the exploration of physics beyond the Standard Model with CP violation.

\subsection{Null results and new physics: spontaneously broken CP?}

If EDM experiments continue producing null results even as they attain orders of magnitude more sensitivity than ACME (proposals include e.g.~\cite{Kozyryev:2017cwq, Vutha:2017pej, Lim:2018kkn}), theorists must decide whether to doubt that {\em any} new physics near the TeV scale interacts with the Standard Model. Our naive expectation is that anywhere a CP phase is allowed, it should be order one. The CKM phase is order one; the QCD theta angle is not, but we know a simple dynamical mechanism for relaxing it to zero (the axion), unlike generic phases. Furthermore, there are tentative indications from neutrino experiments (so far at low statistical significance) that the CP violating phase in the PMNS matrix is large as well (see e.g.~\cite{Esteban:2016qun}).

Is our intuition that new physics should come with order-one CP violating phases robust, or could there be fundamental reasons (apart from fine tuning) why the CP phases associated with new physics could all be small? One possible explanation for small phases lies in spontaneously broken CP. It is likely that, at a fundamental level, CP is a gauge symmetry; in this case, the CP violation that we see in nature is a result of spontaneous breaking by the VEVs of various scalar fields \cite{Dine:1992ya, Choi:1992xp}. If the fields spontaneously violating CP also violate other symmetries, then their contributions may generically be suppressed by small symmetry-breaking order parameters \cite{Nir:1996am}. For instance, if only flavon VEVs have CP-violating phases, they can effectively contribute a large phase in the CKM matrix when added to other VEVs violating the same flavor symmetries, but they will have subleading contributions to parameters that do not violate flavor. Since \cite{Nir:1996am} focused on the quark sector, it could be interesting to revisit such models given that the neutrino Jarlskog invariant appears to be $\sim 0.03$, much larger than $3 \times 10^{-5}$ in the quark sector \cite{Esteban:2016qun}. 

More generally, UV complete theories could have additional structure suppressing some CP violating effects. It has been observed that ``mirror mediation'' of SUSY breaking, with flavor structure arising from complex structure moduli and SUSY breaking from K\"ahler moduli (or vice versa), suppresses CP phases in soft SUSY breaking terms \cite{Conlon:2007dw}. The structure appears somewhat ad hoc in low energy effective field theory, but arises from a higher dimensional theory with extended SUSY. Relatively little exploration has been carried out of the sizes of small CP-violating phases arising from corrections to this picture. (For a different moduli mediation scenario, see \cite{Ellis:2014tea}, which predicts an electron EDM of about $5 \times 10^{-30}\, e\, {\rm cm}$---not far below the current bound!) 

Detailed model building of the origin of the CKM matrix and complex phases in supersymmetric theories, and their correlation with the predicted size of EDMs, has somewhat fallen out of fashion. We believe that the current rapid improvement in experimental results makes it very timely to revisit these questions: the answer could have major implications for the plausibility of scenarios like mini-split SUSY in light of data.

\subsection{An EDM would reify the hierarchy problem}

If a nonzero electron EDM is detected in the foreseeable future, it will necessarily indicate physics beyond the Standard Model. As we have seen, this physics could arise over a wide range of mass scales. In some models, the particles generating the EDM would likely lie within reach, if not of the LHC, at least of a conceivable future collider. In other cases, they would not. For instance, in \S\ref{sec:oneloopSUSY} we have seen that EDMs arising at one loop in a SUSY theory could come from sleptons with masses approaching 10 TeV, well out of reach of any proposed collider. Scenarios where the electron Yukawa coupling is generated radiatively could come from even higher energy physics, as discussed in \S\ref{sec:bigpicture}. While an EDM discovery would be a solid indicator that there is physics beyond the Standard Model at energies far below the GUT scale, it would not immediately give rise to a ``no-lose'' theorem for technologically feasible colliders.

Despite the lack of a clean no-lose theorem, an electron EDM would clearly motivate renewed enthusiasm for searching for heavy particles. In particular, new CP phases need not be order one, and as we have just discussed, in some models there could be compelling reasons for phases to be small and for the associated EDM-generating particles to be lighter. However, there is another argument for searching for new physics at colliders if EDMs are generated by heavy particles: the Higgs mass fine-tuning problem, which would assume a new and very concrete form.

\begin{figure}[h]
\begin{center}
  \includegraphics[width=0.5\textwidth]{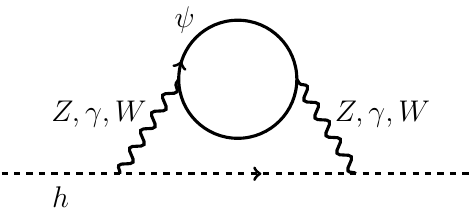}
\end{center}
\caption{Two-loop correction to the Higgs boson mass-squared parameter proportional to the mass squared of any heavy field $\psi$ (here depicted as a fermion for concreteness) with electroweak quantum numbers. Such contributions produce a very concrete, calculable variation on the Higgs fine-tuning problem if the particles generating an EDM are much heavier than the TeV scale.}
\label{fig:hierarchy}
\end{figure}

Discussions of fine-tuning of the Higgs mass are often phrased in terms of quadratic divergences: if we cut off loops of Standard Model particles at a scale $\Lambda$, we obtain corrections to the Higgs mass squared parameter proportional to $\Lambda^2$. However, as is often pointed out by skeptics, UV cutoffs are theorists' conventions; what we really should mean by a hierarchy problem is sensitivity to {\em physical} mass scales, such as masses of heavy particles beyond the Standard Model. To be confident that the hierarchy problem is a problem, we must know that there {\em is} new physics at energies above the weak scale. While there is a compelling argument to be made that the existence of gravity necessitates such physics, one can (and many do) question this logic. However, if we have actual evidence of new physics at high energies from the measurement of a small coefficient for a higher dimension operator like an EDM, it becomes much harder to dismiss the hierarchy problem.

\begin{figure}[!t]
  \centering
  \includegraphics[width=0.6\textwidth]{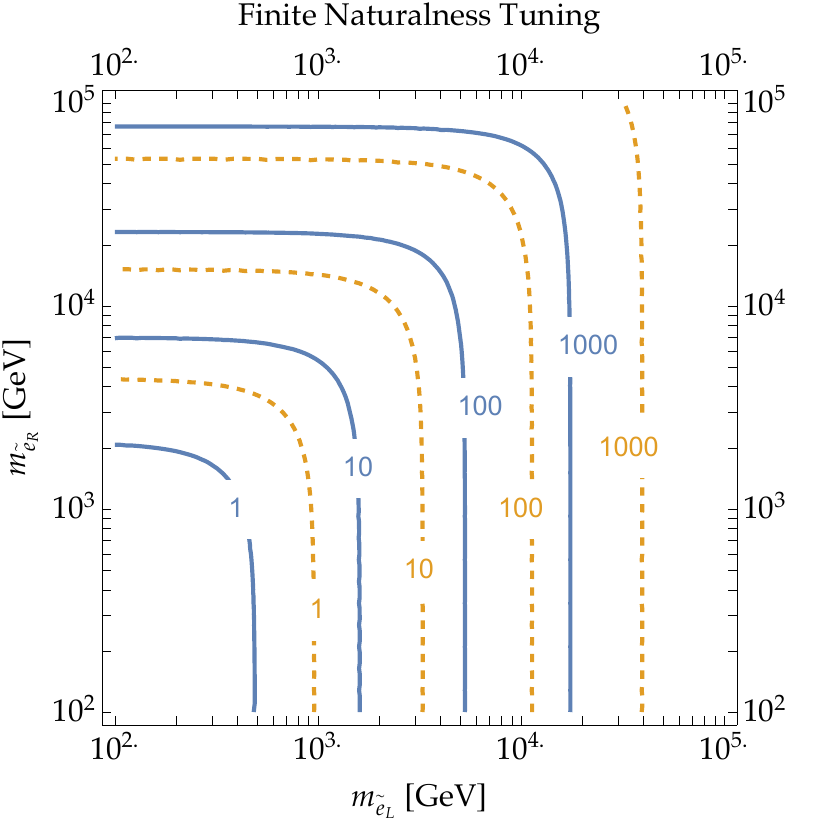}
  \caption{``Finite naturalness'' tuning measure for a model consisting of a bino and left- and right-handed selectrons. Contours correspond to $\Delta = |\delta m^2_{\rm loop} / m_{\rm higgs}^2|$ for two different choices of UV cutoff, $\Lambda = 2 \times 10^{16}~{\rm GeV}$ (blue) and $10^{10}~{\rm GeV}$ (orange), with log divergences included but no power divergences.}
  \label{fig:tuningbinoselectron}
\end{figure}

If heavy particles with electroweak quantum numbers exist, the hierarchy problem can assume a concrete, calculable form in which the Higgs boson mass parameter receives corrections proportional to the physical masses of those new particles. Such dependence always arises from two-loop diagrams as shown in Fig.~\ref{fig:hierarchy}. The size of this loop correction relative to the measured Higgs boson mass, $\Delta \equiv | \delta m^2_{\rm loop} / m_{\rm higgs}^2|$, is a very conservative measure of fine tuning (see e.g.~\cite{Vissani:1997ys, Farina:2013mla, deGouvea:2014xba} for related discussions). The two-loop Higgs mass squared corrections proportional to the mass squared of new fermions or scalars transforming in any representation of the electroweak gauge group have been given in \cite{Farina:2013mla}. (These have logarithmic sensitivity to the UV cutoff, but no power law sensitivity.) Using these expressions, we have plotted the corresponding ``finite naturalness'' tuning contours in Fig.~\ref{fig:tuningbinoselectron} for a model with an EDM generated by particles with the quantum numbers of a bino, left-handed selectron, and right-handed selectron. We see that as the masses approach the 10 TeV scale, the theory is tuned at worse than a percent level. We emphasize that we are {\em not} assuming that the underlying theory is SUSY or that any new particles exist except for the ones that produce the EDM. Already, this minimal model for EDM generation would imply a sharp form of the hierarchy problem, which would be much more difficult for skeptics to dismiss than the usual formulation of the problem. Nonetheless, this problem could only be solved by invoking the same types of physics that solve the usual hierarchy problem, such as supersymmetry or compositeness of the Higgs boson. These in turn imply new particles (such as top partners) whose masses the Higgs is sensitive to, and all of the usual arguments for expecting such particles near the TeV scale would go through---but resting on a firmer foundation.

In summary, we believe that any future nonzero electron EDM measurement would have profound and exciting implications for particle physics. It would immediately provide a much stronger case for pursuing new high energy colliders, and would guarantee that we have more to learn about the fundamental laws of nature. In the meantime, null results of EDM experiments like the recent ACME result provide stringent constraints on theories of new physics and motivate further work to assess what principles could lead to small CP-violating phases if new TeV-scale physics exists. We eagerly await further results from precision measurements.

\section*{Note added}

Shortly after the preprint of this paper was posted to arXiv, an overlapping analysis of the implications of the ACME result appeared in \cite{Panico:2018hal}. Their study includes the effect of operators that mix into the electron EDM at two loops. Some of these operators generate the QULE operator at one loop, and can themselves be generated at tree level, e.g.~via leptoquark exchange. Such EDMs are similar to the two-loop EDMs we discuss generated by QULE box diagrams. However, we have assumed that all particles in the box are BSM particles, while if the operators mixing into the EDM at two loops in \cite{Panico:2018hal} arise at tree level, this would correspond to a box diagram where only one internal line is a BSM particle. Thus, neither our results nor those of \cite{Panico:2018hal} entirely subsume the other.

\section*{Acknowledgments}

We thank John Doyle, Cris Panda, David Pinner and Scott Thomas for discussions. YN is supported by the DOE grant DE-SC0010008. AP is supported in part by an NSF Graduate Research Fellowship grant DGE1745303. MR is supported in part by the NASA ATP Grant NNX16AI12G. CC, QL, and MR are supported in part by the DOE Grant DE-SC0013607. 

\appendix

\section{Classification of QULE box diagrams}
\label{app:QULE}

\renewcommand{\arraystretch}{1.3}

This appendix lists all possible quantum numbers of the new vectorlike fermions and complex scalars we introduced to generate $(q\cdot\ell)(\bar{u}\bar{e})$ or $(q\bar{e})\cdot(\ell\bar{u})$ operators through a box diagram. We will make the simplifying assumption that the particles are either (anti-)fundamental or singlet under $SU(3)$ and $SU(2)$. Even with this assumption, there are infinitely many possiblilities as a function of a free parameter $Y$, the hypercharge of one of the particles. Two different criteria are then used to constrain the value of $Y$. The first is to make at least one of the particles an electrically-neutral color singlet, such that all particles can decay to Standard Model particles and the neutral particle. The second is to make at least one of particles couple to a pair of Standard Model particles, such that all particles can decay back to the Standard Model.

For every set of quantum numbers, we also check for three potentially problematic behaviors. The first is whether the particles also generate $(q\cdot\ell)(\bar{u}\bar{e})$ or $(q\bar{e})\cdot(\ell\bar{u})$ operators at tree-level, thus an electron EDM at 1-loop. The second is whether the particles generate $C_S$ operator at tree-level, which can be more dominant than the two-loop electron EDM even after applying the suppression factors in \eqref{eq:de_to_Cs}.  The third is whether the particles cause proton decay. Tree-level proton decay can be caused by scalar particles alone, while loop-level proton decay can happen when the fermions and scalars also generate, through a box diagram, the 4-fermion operators that lead to proton decay. However, no case in this appendix has been found where the particles cause loop-level proton decay without causing tree-level proton decay. Therefore, from this point on, ``proton decay'' will always refer to a tree-level process.

Finally, we note that, as discussed around Fig.~\ref{fig:feynmanQULE}, in some models a coupling of $\psi_1 \psi_2$ to the Higgs boson is allowed, which can generate a 1-loop EDM. This is always true for the models in \S\ref{subsec:qlue} and \S\ref{subsec:qeul} below. On the other hand, the models in \S\ref{subsec:qleu} and \S\ref{subsec:qelu} have an intermediate state $\psi_1 \psi_2$ with the quantum numbers of ${\bar \ell} u$ or $\ell q$, and as such do not generate 1-loop contributions to the EDM in the same way. Furthermore, because there is no possibility for Higgs couplings for either the two-fermion or two-scalar intermediate states obtained by cutting these diagrams, these models do not lead to Barr-Zee contributions. Hence, these are the models in which (unless otherwise flagged in the tables below) we expect the 2-loop EDM generated from the QULE operator to dominate.

For clarity, we list below the quantum numbers of the scalar particle that can cause one or more of the problematic behaviors. The quantum number of scalar mediators for tree-level $(q\cdot\ell)(\bar{u}\bar{e})$ or $(q\bar{e})\cdot(\ell\bar{u})$ were discussed in \S\ref{sec:QULE} and for proton decay were discussed in \cite{Arnold:2012sd}. For tree-level $C_S$, it is equivalent to finding scalar mediators that generate $(\bar{d}q)\cdot(\bar{\ell}e)$, $(q\cdot\ell)(\bar{u}\bar{e})$, $(q\bar{e})\cdot(\ell\bar{u})$, or  $(q\bar{u})\cdot(\ell\bar{e})$, as discussed in \S\ref{subsec:RGE_qule} and at the beginning of \S\ref{sec:QULE}.
\begin{center}
\begin{tabular}{c | c | c | c}
    \hline
    Quantum number of scalar & EDM at 1-loop? & Tree-level $C_S$? & Proton decay?\\
    \hline
    $(3, 3, -1/3)$ & No & No & Yes \\
    \hline
    $(3, 2, 7/6)$ & Yes & Yes & No \\
    \hline
    $(3, 2, 1/6)$ & No & No & Yes \\
    \hline
    $(3, 2, -5/6)$ & No & No & Yes \\
    \hline
    $(3, 1, -1/3)$ & Yes & Yes & Yes \\
    \hline
    $(3, 1, -4/3)$ & No & No & Yes \\
    \hline
    $(1, 2, 1/2)$ & No & Yes & No \\
    \hline
    $(1, 2, -1/2)$ & No & Yes & No \\
    \hline
\end{tabular}
\end{center}
We also list here all possible renormalizable couplings to Standard Model particles for each quantum number pattern that appears in the box diagram. Couplings which only differ by exchanging $\phi$($\psi$) and $\phi^{\dagger}$($\bar{\psi}$) are considered distinct, because in the box diagram, the hypercharge of $\phi$($\psi$) is in general a linear equation in the hypercharge of other particles, and choosing $\phi$($\psi$) or $\phi^{\dagger}$($\bar{\psi}$) couplings will give different hypercharge value.

\begin{center}
\begin{tabular}{ C{2cm} | C{3cm} | C{2cm} | C{3cm} | c }
    \hline
    $\phi$ charge & \multicolumn{2}{c|}{coupling and $Y$ value} & \multicolumn{2}{c}{coupling and $Y$ value} \\
    \hline
    $(3, 2)_Y$ & $\phi\cdot \ell\bar{u}$, $\phi^{\dagger}q\bar{e}$ & $Y = 7/6$ & $\phi\cdot \ell\bar{d}$ & $Y = 1/6$ \\
    \hline
    $(3, 1)_Y$ & $\phi\bar{u}\bar{e}$, $\phi^{\dagger}q\cdot \ell$ & $Y = -1/3$ & $\phi\bar{d}\bar{e}$& $Y = -4/3$ \\
    \hline
    $(1, 2)_Y$ & $\phi\cdot q\bar{u}$, $\phi^{\dagger}q\bar{d}$, $\phi^{\dagger}\ell\bar{e}$ & $Y = 1/2$ & $\phi\cdot  q\bar{d}$, $\phi\cdot \ell\bar{e}$, $\phi^{\dagger}q\bar{u}$& $Y = -1/2$ \\
    \hline
    \multirow{2}{*}{$(1, 1)_Y$} & $\phi \ell\cdot \ell$& $Y = 1$ & $\phi\bar{e}\bar{e}$& $Y = -2$\\
    \cline{2-5}
    & $\phi^{\dagger} \ell\cdot \ell$& $Y = -1$ & $\phi^{\dagger}\bar{e}\bar{e}$& $Y = 2$\\
    \hline
\end{tabular}
\end{center}

\begin{center}
\begin{tabular}{ C{2cm} | C{3cm} | C{2cm} | C{3cm} | c }
    \hline
    $\psi$ charge & \multicolumn{2}{c|}{coupling and $Y$ value} & \multicolumn{2}{c}{coupling and $Y$ value}\\
    \hline
    \multirow{2}{*}{$(3, 2)_Y$} & $h\cdot\psi\bar{u}$, $h^{\dagger}\psi\bar{d}$ & $Y = 1/6$ & $h^{\dagger}\psi\bar{u}$ & $Y = 7/6$ \\
    \cline{2-5}
    & $h\cdot\psi\bar{d}$ & $Y = -5/6$ & N/A & N/A\\
    \hline
    $(3, 1)_Y$ & $q\cdot h\bar{\psi}$& $Y = 2/3$ & $h^{\dagger}q\bar{\psi}$& $Y = -1/3$\\
    \hline
    \multirow{2}{*}{$(1, 2)_Y$} & $h\cdot\psi\bar{e}$& $Y = -3/2$ & $h^{\dagger}\psi\bar{e}$& $Y = -1/2$\\
    \cline{2-5}
    & $h\cdot\bar{\psi}\bar{e}$& $Y = 3/2$ & $h^{\dagger}\bar{\psi}\bar{e}$& $Y = 1/2$\\
    \hline
    $(1, 1)_Y$ & $h^{\dagger}\ell\psi$& $Y = 1$ & $h^{\dagger}\ell\bar{\psi}$& $Y = -1$\\
    \hline
    
\end{tabular}
\end{center}

\newpage
\begin{landscape}

\subsection{$(q\cdot\ell)(\bar{e}\bar{u})$}\label{subsec:qleu}
The box diagram that generates this operator is
\begin{center}
\begin{tikzpicture}[line width=1.5 pt]
\draw[fermion] (-4.5,0)--(-1.5,0);
\node at (-3,-0.5) {$\bar{e}\ (1, 1)_{1}$};
\draw[fermion] (4.5,0)--(1.5,0);
\node at (3,-0.5) {$\bar{u}\ (\bar{3}, 1)_{-2/3}$};
\draw[fermion] (-1.5,0)--(1.5,0);
\node at (0, -0.5) {$\psi_2$};

\draw[scalar] (-1.5, 3) -- (-1.5, 0);
\node at (-2, 1.5) {$\phi_1$};
\draw[scalar] (1.5, 3)  -- (1.5, 0);
\node at (2, 1.5) {$\phi_2$};

\draw[fermion] (-4.5,3)--(-1.5,3);
\node at (-3,3.5) {$q\ (3, 2)_{1/6}$};
\draw[fermion] (4.5,3)--(1.5,3);
\node at (3,3.5) {$\ell\ (1, 2)_{-1/2}$};
\draw[fermion] (-1.5,3)--(1.5,3);
\node at (0, 3.5) {$\psi_1$};
\end{tikzpicture}
\end{center}
The most general quantum numbers for this diagram are:
\begin{center}
  \begin{tabular}{ c | c | c | c}
    \hline
    $\phi_1$ & $\phi_2$ & $\psi_1$ & $\psi_2$ \\ \hline
    $(3, 2)_Y$ & $(1, 2)_{-1/3-Y}$ & $(1, 1)_{1/6-Y}$ & $(3, 2)_{1+Y}$\\ \hline
    $(3, 1)_Y$ & $(1, 1)_{-1/3-Y}$ & $(1, 2)_{1/6-Y}$ & $(3, 1)_{1+Y}$\\ \hline
    $(1, 1)_Y$ & $(3, 1)_{-1/3-Y}$ & $(3, 2)_{1/6-Y}$ & $(1, 1)_{1+Y}$\\ \hline
    $(1, 2)_Y$ & $(3, 2)_{-1/3-Y}$ & $(3, 1)_{1/6-Y}$ & $(1, 2)_{1+Y}$\\
    \hline
  \end{tabular}
\end{center}

\subsubsection{$Y$ assignment that makes at least one particle electrically-neutral and color singlet} \label{sec:A11tab}
\begin{center}
  \begin{tabular}{ c | c | c | c | c | c | c | c}
    \hline
    $Y$ &$\phi_1$ & $\phi_2$ & $\psi_1$ & $\psi_2$ & EDM at 1-loop? &Tree-level $C_S$? &  Proton decay? \\ \hline
    $1/6$ & $(3, 2)_{1/6}$ & $(1, 2)_{-1/2}$ & $(1, 1)_{0}$ & $(3, 2)_{7/6}$ & No & Yes & Yes \\ \hline
    $-1/3$ & ${(3, 1)_{-1/3}}$ & $ (1, 1)_{0}$ & ${(1, 2)_{1/2}}$ & $(3, 1)_{2/3}$ & Yes, $(q\cdot\ell)(\bar{u}\bar{e})$ & Yes & Yes\\ \hline
    $2/3$ & $(3, 1)_{2/3}$ & $(1, 1)_{-1}$ & ${(1, 2)_{-1/2}}$ & $(3, 1)_{5/3}$ & No & No & No \\ \hline
    $0$ & ${(1, 1)_{0}}$ & ${(3, 1)_{-1/3}}$ & $(3, 2)_{1/6}$ & $(1, 1)_{1}$ & Yes, $(q\cdot\ell)(\bar{u}\bar{e})$& Yes & Yes\\ \hline
    $-1$ & $(1, 1)_{-1}$ & $(3, 1)_{2/3}$ & $(3, 2)_{7/6}$ & ${(1, 1)_{0}}$ & No & No & No \\ \hline
    $1/2$ & ${(1, 2)_{1/2}}$ & $(3, 2)_{-5/6}$ & $(3, 1)_{-1/3}$ & $(1, 2)_{3/2}$ & No & Yes & Yes\\ \hline
    $-1/2$ & ${(1, 2)_{-1/2}}$ & $(3, 2)_{1/6}$ & $(3, 1)_{2/3}$ & ${(1, 2)_{1/2}}$ & No & Yes & Yes \\ \hline
    $-3/2$ & $(1, 2)_{-3/2}$ & $(3, 2)_{7/6}$ & $(3, 1)_{5/3}$ & ${(1, 2)_{-1/2}}$ & Yes, $(q\bar{e})\cdot(\ell\bar{u})$ & Yes &  No \\ \hline
  \end{tabular} 
\end{center}

\subsubsection{$Y$ assignment that couples at least one particle to a pair of Standard Model particles}
For this section, each row of the general table is given a separate table.

\begin{center}
\begin{tabular}{ C{1.1cm} | C{2.675cm} | C{2.675cm} | C{2.675cm} | C{2.675cm} | c | c | c }
    \hline
    \multicolumn{8}{c}{$(q\cdot\ell)(\bar{e}\bar{u})$ Row 1}\\
    \hline
    \multirow{2}{*}{$Y$} & \multicolumn{4}{c|}{SM coupling} & \multirow{2}{*}{EDM at 1-loop?} & \multirow{2}{*}{Tree-level $C_S$?} & \multirow{2}{*}{Proton decay?}\\
    \cline{2-5}
    & $\phi_1$ $(3,2)_Y$ & $\phi_2$ $(1, 2)_{-1/3-Y}$ & $\psi_1$ $(1, 1)_{1/6-Y}$ & $\psi_2$ $(3, 2)_{1+Y}$ &&&\\
    \hline
    $-5/6$ & N/A & $\phi\cdot q\bar{u}$, $\phi^{\dagger}q\bar{d}$, $\phi^{\dagger}\ell\bar{e}$ & $h^{\dagger}\ell\psi$ & $h\cdot\psi\bar{u}$, $h^{\dagger}\psi\bar{d}$ & No & Yes & Yes\\
    \hline
    $1/6$ & $\phi\cdot \ell\bar{d}$ & $\phi\cdot q\bar{d}$, $\phi\cdot \ell\bar{e}$, $\phi^{\dagger} q\bar{u}$ & N/A & $h^{\dagger}\psi\bar{u}$ & No & Yes & Yes\\
    \hline
    $7/6$ & $\phi\cdot \ell\bar{u}$, $\phi^{\dagger}q\bar{e}$ & N/A & $h^{\dagger}\ell\bar{\psi}$ & N/A &Yes, $(q\bar{e})\cdot(\ell\bar{u})$ & Yes & No \\
    \hline
    $-11/6$ & N/A & N/A & N/A & $h\cdot\psi\bar{d}$ & No & No & No\\
    \hline
\end{tabular}
\end{center}

\begin{center}
\begin{tabular}{ C{1cm} | C{2.7cm} | C{2.7cm} | C{2.7cm} | C{2.7cm} | c | c | c }
    \hline
    \multicolumn{8}{c}{$(q\cdot\ell)(\bar{e}\bar{u})$ Row 2}\\
    \hline
    \multirow{2}{*}{$Y$} & \multicolumn{4}{c|}{SM coupling} & \multirow{2}{*}{EDM at 1-loop?} & \multirow{2}{*}{Tree-level $C_S$?} & \multirow{2}{*}{Proton decay?}\\
    \cline{2-5}
    & $\phi_1$ $(3, 1)_Y$ & $\phi_2$ $(1, 1)_{-1/3-Y}$ & $\psi_1$ $(1, 2)_{1/6-Y}$ & $\psi_2$ $(3, 1)_{1+Y}$ &&&\\
    \hline
    $-1/3$ & $\phi\bar{u}\bar{e}$, $\phi^{\dagger} q\cdot \ell$ & N/A & $h^{\dagger}\bar{\psi}\bar{e}$ & $q\cdot h\bar{\psi}$ & Yes, $(q\cdot\ell)(\bar{u}\bar{e})$ & Yes & Yes\\
    \hline
    $-4/3$ & $\phi\bar{d}\bar{e}$ & $\phi \ell\cdot \ell$ & $h\cdot\bar{\psi}\bar{e}$ & $h^{\dagger}q\bar{\psi}$ & No & No & Yes\\
    \hline
    $5/3$ & N/A & $\phi\bar{e}\bar{e}$ & $h\cdot\psi\bar{e}$ & N/A & No & No & No\\
    \hline
    $2/3$ & N/A & $\phi^{\dagger} \ell\cdot \ell$ & $h^{\dagger}\psi\bar{e}$ & N/A & No & No & No \\
    \hline
    $-7/3$ & N/A & $\phi^{\dagger}\bar{e}\bar{e}$ & N/A & N/A & No & No & No\\
    \hline
\end{tabular}
\end{center}

\begin{center}
\begin{tabular}{ C{1cm} | C{2.7cm} | C{2.7cm} | C{2.7cm} | C{2.7cm} | c | c | c }
    \hline
    \multicolumn{8}{c}{$(q\cdot\ell)(\bar{e}\bar{u})$ Row 3}\\
    \hline
    \multirow{2}{*}{$Y$} & \multicolumn{4}{c|}{SM coupling} & \multirow{2}{*}{EDM at 1-loop?} & \multirow{2}{*}{Tree-level $C_S$?} & \multirow{2}{*}{Proton decay?}\\
    \cline{2-5}
    & $\phi_1$ $(1, 1)_{Y}$ & $\phi_2$ $(3, 1)_{-1/3-Y}$ & $\psi_1$ $(3, 2)_{1/6-Y}$ & $\psi_2$ $(1, 1)_{1+Y}$ &&&\\
    \hline
    $1$ & $\phi \ell\cdot \ell$ & $\phi\bar{d}\bar{e}$ & $h\cdot\psi\bar{d}$ & N/A & No & No & Yes \\
    \hline
    $0$ & N/A & $\phi\bar{u}\bar{e}$, $\phi^{\dagger} q\cdot \ell$ & $h\cdot\psi\bar{u}$, $h^{\dagger}\psi\bar{d}$ & $h^{\dagger}\ell\psi$ & Yes, $(q\cdot\ell)(\bar{u}\bar{e})$ & Yes & Yes \\
    \hline
    $-1$ & $\phi^{\dagger} \ell\cdot \ell$ & N/A & $h^{\dagger}\psi\bar{u}$ & N/A & No & No & No\\
    \hline
    $-2$ & $\phi\bar{e}\bar{e}$ & N/A & N/A & $h^{\dagger}\ell\bar{\psi}$ & No & No & No\\
    \hline
    $2$ & $\phi^{\dagger}\bar{e}\bar{e}$ & N/A & N/A & N/A & No & No & No\\
    \hline
\end{tabular}
\end{center}

\begin{center}
\begin{tabular}{ C{1cm} | C{2.7cm} | C{2.7cm} | C{2.7cm} | C{2.7cm} | c | c | c }
    \hline
    \multicolumn{8}{c}{$(q\cdot\ell)(\bar{e}\bar{u})$ Row 4}\\
    \hline
    \multirow{2}{*}{$Y$} & \multicolumn{4}{c|}{SM coupling} & \multirow{2}{*}{EDM at 1-loop?} & \multirow{2}{*}{Tree-level $C_S$?} & \multirow{2}{*}{Proton decay?}\\
    \cline{2-5}
    & $\phi_1$ $(1, 2)_Y$ & $\phi_2$ $(3, 2)_{-1/3-Y}$ & $\psi_1$ $(3, 1)_{1/6-Y}$ & $\psi_2$ $(1, 2)_{1+Y}$ &&&\\
    \hline
    $1/2$ & $\phi\cdot q\bar{u}$, $\phi^{\dagger} q \bar{d}$, $\phi^{\dagger}\ell\bar{e}$ & N/A & $h^{\dagger}q\bar{\psi}$ & $h\cdot\bar{\psi}\bar{e}$ & No & Yes & Yes \\
    \hline
    $-1/2$ & $\phi\cdot q\bar{d}$, $\phi\cdot \ell\bar{e}$, $\phi^{\dagger} q\bar{u}$ & $\phi\cdot \ell\bar{d}$ & $q\cdot h\bar{\psi}$ & $h^{\dagger}\bar{\psi}\bar{e}$A & No & Yes & Yes\\
    \hline
    $-3/2$ & N/A & $\phi\cdot \ell\bar{u}$, $\phi^{\dagger}q\bar{e}$ & N/A & $h^{\dagger}\psi\bar{e}$ & Yes, $(q\cdot\ell)(\bar{u}\bar{e})$ & Yes & No \\
    \hline
    $-5/2$ & N/A & N/A & N/A & $h\cdot\psi\bar{e}$ & No & No & No \\
    \hline
\end{tabular}
\end{center}

\subsection{$(q\cdot\ell)(\bar{u}\bar{e})$} \label{subsec:qlue}
$(q\cdot\ell)(\bar{u}\bar{e})$ is equivalent to $(q\cdot\ell)(\bar{e}\bar{u})$ operator in \S\ref{subsec:qleu}, except here we are considering a different diagram:
\begin{center}
\begin{tikzpicture}[line width=1.5 pt]
\draw[fermion] (-4.5,0)--(-1.5,0);
\node at (-3,-0.5) {$\bar{u}\ (\bar{3}, 1)_{-2/3}$};
\draw[fermion] (4.5,0)--(1.5,0);
\node at (3,-0.5) {$\bar{e}\ (\bar{1}, 1)_{1}$};
\draw[fermion] (-1.5,0)--(1.5,0);
\node at (0, -0.5) {$\psi_2$};

\draw[scalar] (-1.5, 3) -- (-1.5, 0);
\node at (-2, 1.5) {$\phi_1$};
\draw[scalar] (1.5, 3)  -- (1.5, 0);
\node at (2, 1.5) {$\phi_2$};

\draw[fermion] (-4.5,3)--(-1.5,3);
\node at (-3,3.5) {$q\ (3, 2)_{1/6}$};
\draw[fermion] (4.5,3)--(1.5,3);
\node at (3,3.5) {$\ell\ (1, 2)_{-1/2}$};
\draw[fermion] (-1.5,3)--(1.5,3);
\node at (0, 3.5) {$\psi_1$};
\end{tikzpicture}
\end{center}
This case is somewhat less interesting, in that the quantum numbers allow for a Yukawa coupling of $\psi_1 \psi_2$ to the Higgs boson and hence a 1-loop EDM.
The most general quantum numbers for this diagram are:
\begin{center}
\begin{tabular}{ c | c | c | c}
    \hline
    $\phi_1$ & $\phi_2$ & $\psi_1$ & $\psi_2$ \\ \hline
    $(3, 2)_Y$ & $(1, 2)_{-1/3-Y}$ & $(1, 1)_{1/6-Y}$ & $(1, 2)_{Y-2/3}$\\ \hline
    $(3, 1)_Y$ & $(1, 1)_{-1/3-Y}$ & $(1, 2)_{1/6-Y}$ & $(1, 1)_{Y-2/3}$\\ \hline
    $(1, 1)_{Y}$ & $(3, 1)_{-1/3-Y}$ & $(3, 2)_{1/6-Y}$ & $(\bar{3}, 1)_{Y-2/3}$\\ \hline
    $(1, 2)_{Y}$ & $(3, 2)_{-1/3-Y}$ & $(3, 1)_{1/6-Y}$ & $(\bar{3}, 2)_{Y-2/3}$\\
    \hline
\end{tabular}
\end{center}
\subsubsection{$Y$ assignment that makes at least one particle electrically-neutral and color singlet}
\begin{center}
  \begin{tabular}{ c | c | c | c | c | c | c | c}
    \hline
    $Y$ &$\phi_1$ & $\phi_2$ & $\psi_1$ & $\psi_2$ & EDM at 1-loop? & Tree-level $C_S$? & Proton decay? \\ \hline
    $1/6$ & $(3, 2)_{1/6}$ & $(1, 2)_{-1/2}$ & $(1, 1)_{0}$ & $(1, 2)_{-1/2}$ & No & Yes & Yes\\ \hline
    $-1/3$ & ${(3, 1)_{-1/3}}$ & ${(1, 1)_{0}}$ & ${(1, 2)_{1/2}}$ & $(1, 1)_{-1}$ &Yes, $(q\cdot\ell)(\bar{u}\bar{e})$& Yes & Yes\\ \hline
    $2/3$ & $(3, 1)_{2/3}$ & $(1, 1)_{-1}$ & ${(1, 2)_{-1/2}}$ & $(1, 1)_{0}$ & No & No & No \\ \hline
    $0$ & ${(1, 1)_{0}}$ & ${(3, 1)_{-1/3}}$ & $(3, 2)_{1/6}$ & $(\bar{3}, 1)_{-2/3}$ & Yes, $(q\cdot\ell)(\bar{u}\bar{e})$ &Yes& Yes\\ \hline
    $1/2$ & ${(1, 2)_{1/2}}$ & $(3, 2)_{-5/6}$ & $(3, 1)_{-1/3}$ & $(\bar{3}, 2)_{-1/6}$ & No & Yes & Yes\\ \hline
    $-1/2$ & ${(1, 2)_{-1/2}}$ & $(3, 2)_{1/6}$ & $(3, 1)_{2/3}$ & $(\bar{3}, 2)_{-7/6}$ & No & Yes & Yes \\ \hline
  \end{tabular}
\end{center}

\subsubsection{$Y$ assignment that couples at least one particle to a pair of Standard Model particles}
For this section, each row of the general table is given a separate table.

\begin{center}
\begin{tabular}{ C{1cm} | C{2.7cm} | C{2.7cm} | C{2.7cm} | C{2.7cm} | c | c | c }
    \hline
    \multicolumn{8}{c}{$(q\cdot\ell)(\bar{u}\bar{e})$ Row 1}\\
    \hline
    \multirow{2}{*}{$Y$} & \multicolumn{4}{c|}{SM coupling} & \multirow{2}{*}{EDM at 1-loop?} & \multirow{2}{*}{Tree-level $C_S$?} & \multirow{2}{*}{Proton decay?}\\
    \cline{2-5}
    & $\phi_1$ $(3,2)_Y$ & $\phi_2$ $(1, 2)_{-1/3-Y}$ & $\psi_1$ $(1, 1)_{1/6-Y}$ & $\psi_2$ $(1, 2)_{Y-2/3}$ &&&\\
    \hline
    $-5/6$ & N/A & $\phi\cdot q\bar{u}$, $\phi^{\dagger}q\bar{d}$, $\phi^{\dagger}\ell\bar{e}$ & $h^{\dagger}\ell\psi$ & $h\cdot\psi\bar{e}$ & No & Yes & Yes \\
    \hline
    $1/6$ & $\phi\cdot \ell\bar{d}$ & $\phi\cdot q\bar{d}$, $\phi\cdot \ell\bar{e}$, $\phi^{\dagger} q\bar{u}$ & N/A & $h^{\dagger}\psi\bar{e}$ & No & Yes & Yes \\
    \hline
    $7/6$ & $\phi\cdot \ell\bar{u}$, $\phi^{\dagger}q\bar{e}$ & N/A & $h^{\dagger}\ell\bar{\psi}$ & $h^{\dagger}\bar{\psi}\bar{e}$ & Yes, $(q\bar{e})\cdot(\ell\bar{u})$ & Yes & No \\
    \hline
    $13/6$ & N/A & N/A & N/A & $h\cdot\bar{\psi}\bar{e}$ & No & No & No\\
    \hline
\end{tabular}
\end{center}

\begin{center}
\begin{tabular}{ C{1cm} | C{2.7cm} | C{2.7cm} | C{2.7cm} | C{2.7cm} | c | c | c }
    \hline
    \multicolumn{8}{c}{$(q\cdot\ell)(\bar{u}\bar{e})$ Row 2}\\
    \hline
    \multirow{2}{*}{$Y$} & \multicolumn{4}{c|}{SM coupling} & \multirow{2}{*}{EDM at 1-loop?} & \multirow{2}{*}{Tree-level $C_S$?} & \multirow{2}{*}{Proton decay?}\\
    \cline{2-5}
    & $\phi_1$ $(3, 1)_Y$ & $\phi_2$ $(1, 1)_{-1/3-Y}$ & $\psi_1$ $(1, 2)_{1/6-Y}$ & $\psi_2$ $(1, 1)_{Y-2/3}$ &&&\\
    \hline
    $-4/3$ & $\phi\bar{d}\bar{e}$ & $\phi \ell\cdot \ell$ & $h\cdot\bar{\psi}\bar{e}$ & N/A & No & No & Yes \\
    \hline
    $-1/3$ & $\phi\bar{u}\bar{e}$, $\phi^{\dagger} q\cdot \ell$ & N/A & $h^{\dagger}\bar{\psi}\bar{e}$ & $h^{\dagger}\ell\bar{\psi}$ & Yes, $(q\cdot\ell)(\bar{u}\bar{e})$ & Yes & Yes \\
    \hline
    $5/3$ & N/A & $\phi\bar{e}\bar{e}$ & $h\cdot\psi\bar{e}$ & $h^{\dagger}\ell\psi$ & No & No & No \\
    \hline
    $2/3$ & N/A & $\phi^{\dagger} \ell\cdot \ell$ & $h^{\dagger}\psi\bar{e}$ & N/A & No & No & No \\
    \hline
    $-7/3$ & N/A & $\phi^{\dagger}\bar{e}\bar{e}$ & N/A & N/A & No & No & No \\
    \hline
\end{tabular}
\end{center}

\begin{center}
\begin{tabular}{ C{1cm} | C{2.7cm} | C{2.7cm} | C{2.7cm} | C{2.7cm} | c | c | c }
    \hline
    \multicolumn{8}{c}{$(q\cdot\ell)(\bar{u}\bar{e})$ Row 3}\\
    \hline
    \multirow{2}{*}{$Y$} & \multicolumn{4}{c|}{SM coupling} & \multirow{2}{*}{EDM at 1-loop?} & \multirow{2}{*}{Tree-level $C_S$?} & \multirow{2}{*}{Proton decay?}\\
    \cline{2-5}
    & $\phi_1$ $(1, 1)_{Y}$ & $\phi_2$ $(3, 1)_{-1/3-Y}$ & $\psi_1$ $(3, 2)_{1/6-Y}$ & $\psi_2$ $(\bar{3}, 1)_{Y-2/3}$ &&&\\
    \hline
    $1$ & $\phi \ell\cdot \ell$ & $\phi\bar{d}\bar{e}$ & $h\cdot\psi\bar{d}$ & $h^{\dagger}q\psi$ & No & No & Yes\\
    \hline
    $0$ & N/A & $\phi\bar{u}\bar{e}$, $\phi^{\dagger} q\cdot \ell$ & $h\cdot\psi\bar{u}$, $h^{\dagger}\psi\bar{d}$ & $q\cdot h\psi$ & Yes, $(q\cdot\ell)(\bar{u}\bar{e})$ & Yes & Yes \\
    \hline
    $-1$ & $\phi^{\dagger} \ell\cdot \ell$ & N/A & $h^{\dagger}\psi\bar{u}$ & N/A & No & No & No \\
    \hline
    $-2$ & $\phi\bar{e}\bar{e}$ & N/A & N/A & N/A & No & No & No\\
    \hline
    $2$ & $\phi^{\dagger}\bar{e}\bar{e}$ & N/A & N/A & N/A & No & No & No \\
    \hline
\end{tabular}
\end{center}

\begin{center}
\begin{tabular}{ C{1cm} | C{2.7cm} | C{2.7cm} | C{2.7cm} | C{2.7cm} | c | c | c }
    \hline
    \multicolumn{8}{c}{$(q\cdot\ell)(\bar{u}\bar{e})$ Row 4}\\
    \hline
    \multirow{2}{*}{$Y$} & \multicolumn{4}{c|}{SM coupling} & \multirow{2}{*}{EDM at 1-loop?} & \multirow{2}{*}{Tree-level $C_S$?} & \multirow{2}{*}{Proton decay?}\\
    \cline{2-5}
    & $\phi_1$ $(1, 2)_Y$ & $\phi_2$ $(3, 2)_{-1/3-Y}$ & $\psi_1$ $(3, 1)_{1/6-Y}$ & $\psi_2$ $(\bar{3}, 2)_{Y-2/3}$ &&&\\
    \hline
    $1/2$ & $\phi\cdot q\bar{u}$, $\phi^{\dagger} q \bar{d}$, $\phi^{\dagger}\ell\bar{e}$ & N/A & $h^{\dagger}q\bar{\psi}$ & $h\cdot\bar{\psi}\bar{u}$, $h^{\dagger}\bar{\psi}\bar{d}$ & No & Yes & Yes \\
    \hline
    $-1/2$ & $\phi\cdot q\bar{d}$, $\phi\cdot \ell\bar{e}$, $\phi^{\dagger} q\bar{u}$ & $\phi\cdot \ell\bar{d}$ & $q\cdot h\bar{\psi}$ & $h^{\dagger}\bar{\psi}\bar{u}$ & No & Yes & Yes\\
    \hline
    $-3/2$ & N/A & $\phi\cdot \ell\bar{u}$, $\phi^{\dagger}q\bar{e}$ & N/A & N/A & Yes, $(q\cdot\ell)(\bar{u}\bar{e})$ & Yes & No \\
    \hline
    $3/2$ & N/A & N/A & N/A & $h\cdot\bar{\psi}\bar{d}$ & No & No & No \\
    \hline
\end{tabular}
\end{center}

\subsection{$(q\bar{e})\cdot(\ell \bar{u})$}\label{subsec:qelu}
The box diagram that generates this operator is
\begin{center}
\begin{tikzpicture}[line width=1.5 pt]
\draw[fermion] (4.5,3)--(1.5,3);
\node at (3,3.5) {$\bar{e}\ (1, 1)_{1}$};

\draw[fermion] (-4.5,0)--(-1.5,0);
\node at (-3,-0.5) {$\ell\ (1, 2)_{-1/2}$};

\draw[fermion] (4.5,0)--(1.5,0);
\node at (3,-0.5) {$\bar{u}\ (\bar{3}, 1)_{-2/3}$};

\draw[scalar] (-1.5, 3) -- (-1.5, 0);
\node at (-2, 1.5) {$\phi_1$};

\draw[fermion] (-1.5,3)--(1.5,3);
\node at (0, 3.5) {$\psi_1$};

\draw[scalar] (1.5, 3)  -- (1.5, 0);
\node at (2, 1.5) {$\phi_2$};

\draw[fermion] (-1.5,0)--(1.5,0);
\node at (0, -0.5) {$\psi_2$};

\draw[fermion] (-4.5,3)--(-1.5,3);
\node at (-3,3.5) {$q\ (3, 2)_{1/6}$};
\end{tikzpicture}
\end{center}
The most general quantum numbers for this diagram are:
\begin{center}
  \begin{tabular}{ c | c | c | c}
    \hline
    $\phi_1$ & $\phi_2$ & $\psi_1$ & $\psi_2$ \\ \hline
    $(1, 1)_{Y}$ & $(3, 2)_{7/6-Y}$ & $(3, 2)_{1/6-Y}$ & $(1, 2)_{Y-1/2}$ \\ \hline
    $(1, 2)_{Y}$ & $(3, 1)_{7/6-Y}$ & $(3, 1)_{1/6-Y}$ & $(1, 1)_{Y-1/2}$\\ \hline
    $(3, 2)_{Y}$ & $(1, 1)_{7/6-Y}$ & $(1, 1)_{1/6-Y}$ & $(3, 1)_{Y-1/2}$ \\ \hline
    $(3, 1)_{Y}$ & $(1, 2)_{7/6-Y}$ & $(1, 2)_{1/6-Y}$ & $(3, 2)_{Y-1/2}$\\
    \hline
  \end{tabular}
\end{center}

\subsubsection{$Y$ assignment that makes at least one particle electrically-neutral and color singlet}
\begin{center}
  \begin{tabular}{ c | c | c | c | c | c | c | c}
    \hline
    $Y$ &$\phi_1$ & $\phi_2$ & $\psi_1$ & $\psi_2$ & EDM at 1-loop? & Tree-level $C_S$? & Proton decay? \\ \hline
    $0$ & ${(1, 1)_{0}}$ & $(3, 2)_{7/6}$ & $(3, 2)_{1/6}$ & $(1, 2)_{-1/2}$ & Yes, $(q\bar{e})\cdot(\ell\bar{u})$ & Yes & No \\ \hline
    $1/2$  & ${(1, 2)_{1/2}}$ & $(3, 1)_{2/3}$ & ${(3, 1)_{-1/3}}$ & ${(1, 1)_{0}}$ & No& Yes & No\\ \hline
    $-1/2$  & ${(1, 2)_{-1/2}}$ & $(3, 1)_{5/3}$ & $(3, 1)_{2/3}$ & $(1, 1)_{-1}$& No & Yes & No \\ \hline
    $1/6$  & $(3, 2)_{1/6}$ & $(1, 1)_{1}$ & ${(1, 1)_{0}}$ & ${(3, 1)_{-1/3}}$ & No & No & No\\ \hline
    $7/6$  & $(3, 2)_{7/6}$ & ${(1, 1)_{0}}$ & $(1, 1)_{-1}$ & $(3, 1)_{2/3}$ & Yes, $(q\bar{e})\cdot(\ell\bar{u})$ & Yes & No \\ \hline
    $-1/3$ & $(3, 1)_{-1/3}$ & $(1, 2)_{3/2}$ & ${(1, 2)_{1/2}}$ & $(3, 2)_{-5/6}$ & No & Yes & Yes \\ \hline
    $2/3$ & $(3, 1)_{2/3}$ & ${(1, 2)_{1/2}}$ & ${(1, 2)_{-1/2}}$ & $(3, 2)_{1/6}$  & No & Yes & No \\ \hline
    $5/3$ & $(3, 1)_{5/3}$ & ${(1, 2)_{-1/2}}$ & $(1, 2)_{-3/2}$ & $(3, 2)_{7/6}$ & No & Yes & No \\ \hline
  \end{tabular}
\end{center}

\subsubsection{$Y$ assignment that couples at least one particle to a pair of Standard Model particles}
For this section, each row of the general table is given a separate table.
\begin{center}
\begin{tabular}{ C{1cm} | C{2.7cm} | C{2.7cm} | C{2.7cm} | C{2.7cm} | c | c | c }
    \hline
    \multicolumn{8}{c}{$(q\bar{e})\cdot(\ell \bar{u})$ Row 1}\\
    \hline
    \multirow{2}{*}{$Y$} & \multicolumn{4}{c|}{SM coupling} & \multirow{2}{*}{EDM at 1-loop?} & \multirow{2}{*}{Tree-level $C_S$?} & \multirow{2}{*}{Proton decay?}\\
    \cline{2-5}
    & $\phi_1$ $(1, 1)_{Y}$ & $\phi_2$ $(3, 2)_{7/6-Y}$ & $\psi_1$ $(3, 2)_{1/6-Y}$ & $\psi_2$ $(1, 2)_{Y-1/2}$ &&&\\
    \hline
    $1$ & $\phi \ell\cdot \ell$ & $\phi\cdot \ell\bar{d}$ & $h\cdot\psi\bar{d}$ & $h^{\dagger}\bar{\psi}\bar{e}$ & No  & No & Yes\\
    \hline
    $0$ & N/A & $\phi\cdot \ell\bar{u}$, $\phi^{\dagger}q\bar{e}$ & $h\cdot\psi\bar{u}$, $h^{\dagger}\psi\bar{d}$ & $h^{\dagger}\psi\bar{e}$ & Yes, $(q\bar{e})\cdot(\ell\bar{u})$ & Yes & No \\
    \hline
    $-1$ & $\phi^{\dagger} \ell\cdot \ell$ & N/A & $h^{\dagger}\psi\bar{u}$  & $h\cdot\psi\bar{e}$ & No & No & No\\
    \hline
    $-2$ & $\phi\bar{e}\bar{e}$ & N/A & N/A & N/A & No & No & No\\
    \hline
    $2$ & $\phi^{\dagger}\bar{e}\bar{e}$ & N/A & N/A & $h\cdot\bar{\psi}\bar{e}$ & No & No & Yes\\
    \hline
\end{tabular}
\end{center}

\begin{center}
\begin{tabular}{ C{1cm} | C{2.7cm} | C{2.7cm} | C{2.7cm} | C{2.7cm} | c | c | c }
    \hline
    \multicolumn{8}{c}{$(q\bar{e})\cdot(\ell \bar{u})$ Row 2}\\
    \hline
    \multirow{2}{*}{$Y$} & \multicolumn{4}{c|}{SM coupling} & \multirow{2}{*}{EDM at 1-loop?} & \multirow{2}{*}{Tree-level $C_S$?} & \multirow{2}{*}{Proton decay?}\\
    \cline{2-5}
    & $\phi_1$ $(1, 2)_Y$ & $\phi_2$ $(3, 1)_{7/6-Y}$ & $\psi_1$ $(3, 1)_{1/6-Y}$ & $\psi_2$ $(1, 1)_{Y-1/2}$ &&&\\
    \hline
    $-1/2$ & $\phi\cdot q\bar{d}$, $\phi\cdot \ell\bar{e}$, $\phi^{\dagger} q\bar{u}$ & N/A & $q\cdot h\bar{\psi}$ & $h^{\dagger}\ell\bar{\psi}$ & No  & Yes & No \\
    \hline
    $1/2$ & $\phi\cdot q\bar{u}$, $\phi^{\dagger} q \bar{d}$, $\phi^{\dagger}\ell\bar{e}$ & N/A & $h^{\dagger}q\bar{\psi}$ & N/A & No  & Yes & No\\
    \hline
    $3/2$ & N/A & $\phi\bar{u}\bar{e}$, $\phi^{\dagger} q\cdot \ell$ & N/A & $h^{\dagger}\ell\psi$ & Yes, $(q\cdot\ell)(\bar{u}\bar{e})$ & Yes & Yes\\
    \hline
    $5/2$ & N/A & $\phi\bar{d}\bar{e}$ & N/A & N/A & No & No & Yes\\
    \hline
\end{tabular}
\end{center}

\begin{center}
\begin{tabular}{ C{1cm} | C{2.7cm} | C{2.7cm} | C{2.7cm} | C{2.7cm} | c | c | c }
    \hline
    \multicolumn{8}{c}{$(q\bar{e})\cdot(\ell \bar{u})$ Row 3}\\
    \hline
    \multirow{2}{*}{$Y$} & \multicolumn{4}{c|}{SM coupling} & \multirow{2}{*}{EDM at 1-loop?} & \multirow{2}{*}{Tree-level $C_S$?} & \multirow{2}{*}{Proton decay?}\\
    \cline{2-5}
    & $\phi_1$ $(3,2)_Y$ & $\phi_2$ $(1, 1)_{7/6-Y}$ & $\psi_1$ $(1, 1)_{1/6-Y}$ & $\psi_2$ $(3, 1)_{Y-1/2}$ &&&\\
    \hline
    $1/6$ & $\phi\cdot \ell\bar{d}$ & $\phi \ell\cdot \ell$ & N/A & $h^{\dagger}q\bar{\psi}$ & No & No & Yes \\
    \hline
    $7/6$ & $\phi\cdot \ell\bar{u}$, $\phi^{\dagger}q\bar{e}$ & N/A & $h^{\dagger}\ell\bar{\psi}$ & $q\cdot h\bar{\psi}$ & Yes, $(q\bar{e})\cdot(\ell\bar{u})$ & Yes & No \\
    \hline
    $-5/6$ & N/A & $\phi^{\dagger}\bar{e}\bar{e}$ & $h^{\dagger}\ell\psi$ & N/A & No  & No & Yes\\
    \hline
    $13/6$ & N/A & $\phi^{\dagger} \ell\cdot \ell$ & N/A & N/A & No & No & No\\
    \hline
    $19/6$ & N/A & $\phi\bar{e}\bar{e}$ & N/A & N/A & No & No & No\\
    \hline
\end{tabular}
\end{center}

\begin{center}
\begin{tabular}{ C{1cm} | C{2.7cm} | C{2.7cm} | C{2.7cm} | C{2.7cm} | c | c | c }
    \hline
    \multicolumn{8}{c}{$(q\bar{e})\cdot(\ell \bar{u})$ Row 4}\\
    \hline
    \multirow{2}{*}{$Y$} & \multicolumn{4}{c|}{SM coupling} & \multirow{2}{*}{EDM at 1-loop?} & \multirow{2}{*}{Tree-level $C_S$?} & \multirow{2}{*}{Proton decay?}\\
    \cline{2-5}
    & $\phi_1$ $(3, 1)_Y$ & $\phi_2$ $(1, 2)_{7/6-Y}$ & $\psi_1$ $(1, 2)_{1/6-Y}$ & $\psi_2$ $(3, 2)_{Y-1/2}$ &&&\\
    \hline
    $-1/3$ & $\phi\bar{u}\bar{e}$, $\phi^{\dagger} q\cdot \ell$ & N/A & $h^{\dagger}\bar{\psi}\bar{e}$ & $h\cdot\psi\bar{d}$ & Yes, $(q\cdot\ell)(\bar{u}\bar{e})$ & Yes & Yes\\
    \hline
    $2/3$ & N/A & $\phi\cdot q\bar{u}$, $\phi^{\dagger}q\bar{d}$, $\phi^{\dagger}\ell\bar{e}$ & $h^{\dagger}\psi\bar{e}$ & $h\cdot\psi\bar{u}$, $h^{\dagger}\psi\bar{d}$ & No & Yes & No\\
    \hline
    $-4/3$ & $\phi\bar{d}\bar{e}$ & N/A & $h\cdot\bar{\psi}\bar{e}$ & N/A & No & No & Yes\\
    \hline
    $5/3$ & N/A & $\phi\cdot q\bar{d}$, $\phi\cdot \ell\bar{e}$, $\phi^{\dagger} q\bar{u}$ & $h\cdot\psi\bar{e}$ & $h^{\dagger}\psi\bar{u}$ & No & Yes & No\\
    \hline
\end{tabular}
\end{center}

\subsection{$(q\bar{e})\cdot(\bar{u}\ell)$}\label{subsec:qeul}
$(q\bar{e})\cdot(\bar{u}\ell)$ is equivalent to $(q\bar{e})\cdot(\ell\bar{u})$ in \S\ref{subsec:qelu}, except here we are considering a different diagram:
\begin{center}
\begin{tikzpicture}[line width=1.5 pt]
\draw[fermion] (4.5,3)--(1.5,3);
\node at (3,3.5) {$\bar{e}\ (1, 1)_{1}$};

\draw[fermion] (-4.5,0)--(-1.5,0);
\node at (-3,-0.5) {$\bar{u}\ (\bar{3}, 1)_{-2/3}$};

\draw[fermion] (4.5,0)--(1.5,0);
\node at (3,-0.5) {$\ell\ (1, 2)_{-1/2}$}; 

\draw[scalar] (-1.5, 3) -- (-1.5, 0);
\node at (-2, 1.5) {$\phi_1$};

\draw[fermion] (-1.5,3)--(1.5,3);
\node at (0, 3.5) {$\psi_1$};

\draw[scalar] (1.5, 3)  -- (1.5, 0);
\node at (2, 1.5) {$\phi_2$};

\draw[fermion] (-1.5,0)--(1.5,0);
\node at (0, -0.5) {$\psi_2$};

\draw[fermion] (-4.5,3)--(-1.5,3);
\node at (-3,3.5) {$q\ (3, 2)_{1/6}$};
\end{tikzpicture}
\end{center}
This case is somewhat less interesting, in that the quantum numbers allow for a Yukawa coupling of $\psi_1 \psi_2$ to the Higgs boson and hence a 1-loop EDM.
The most general quantum numbers for this diagram are:
\begin{center}
  \begin{tabular}{ c | c | c | c}
    \hline
    $\phi_1$ & $\phi_2$ & $\psi_1$ & $\psi_2$ \\ \hline
    $(1, 1)_{Y}$ & $(3, 2)_{7/6-Y}$ & $(3, 2)_{1/6-Y}$ & $(\bar{3}, 1)_{-2/3+Y}$ \\ \hline
    $(1, 2)_{Y}$ & $(3, 1)_{7/6-Y}$ & $(3, 1)_{1/6-Y}$ & $(\bar{3}, 2)_{-2/3+Y}$\\ \hline
    $(3, 2)_{Y}$ & $(1, 1)_{7/6-Y}$ & $(1, 1)_{1/6-Y}$ & $(1, 2)_{-2/3+Y}$ \\ \hline
    $(3, 1)_{Y}$ & $(1, 2)_{7/6-Y}$ & $(1, 2)_{1/6-Y}$ & $(1, 1)_{-2/3+Y}$\\
    \hline
  \end{tabular}
\end{center}

\subsubsection{$Y$ assignment that makes at least one particle electrically-neutral and color singlet} \label{sec:A41tab}
For this section, each row of the general table is given a separate table.

\begin{center}
  \begin{tabular}{ c | c | c | c | c | c | c | c}
    \hline
    $Y$ &$\phi_1$ & $\phi_2$ & $\psi_1$ & $\psi_2$ & EDM at 1-loop? & Tree-level $C_S$? & Proton decay? \\ \hline
    $0$ & ${(1, 1)_{0}}$ & $(3, 2)_{7/6}$ & $(3, 2)_{1/6}$ & $(\bar{3}, 1)_{-2/3}$ & Yes & Yes & No \\ \hline
    $1/2$  & ${(1, 2)_{1/2}}$ & $(3, 1)_{2/3}$ & ${(3, 1)_{-1/3}}$ & $(\bar{3}, 2)_{-1/6}$ & No & Yes & No\\ \hline
    $-1/2$  & ${(1, 2)_{-1/2}}$ & $(3, 1)_{5/3}$ & $(3, 1)_{2/3}$ & $(\bar{3}, 2)_{-7/6}$& No & Yes & No \\ \hline
    $1/6$  & $(3, 2)_{1/6}$ & $(1, 1)_{1}$ & ${(1, 1)_{0}}$ & ${(1, 2)_{-1/2}}$ & No & No & No \\ \hline
    $7/6$  & $(3, 2)_{7/6}$ & ${(1, 1)_{0}}$ & $(1, 1)_{-1}$ & $(1, 2)_{1/2}$ & Yes, $(q\bar{e})\cdot(\ell\bar{u})$ & Yes &No \\ \hline
    $-1/3$ & $(3, 1)_{-1/3}$ & $(1, 2)_{3/2}$ & ${(1, 2)_{1/2}}$ & $(1, 1)_{-1}$ & Yes, $(q\cdot\ell)(\bar{u}\bar{e})$ & Yes & Yes \\ \hline
    $2/3$ & $(3, 1)_{2/3}$ & ${(1, 2)_{1/2}}$ & ${(1, 2)_{-1/2}}$ & $(1, 1)_{0}$ & No & Yes & No \\ \hline
    $5/3$ & $(3, 1)_{5/3}$ & ${(1, 2)_{-1/2}}$ & $(1, 2)_{-3/2}$ & $(1, 1)_{1}$ & No & Yes & No \\ \hline
  \end{tabular}
\end{center}

\subsubsection{$Y$ assignment that couples at least one particle to a pair of Standard Model particles} 
\begin{center}
\begin{tabular}{ C{1cm} | C{2.7cm} | C{2.7cm} | C{2.7cm} | C{2.7cm} | c | c | c }
    \hline
    \multicolumn{8}{c}{$(q\bar{e})\cdot(\bar{u}\ell)$ Row 1}\\
    \hline
    \multirow{2}{*}{$Y$} & \multicolumn{4}{c|}{SM coupling} & \multirow{2}{*}{EDM at 1-loop?} & \multirow{2}{*}{Tree-level $C_S$?} & \multirow{2}{*}{Proton decay?}\\
    \cline{2-5}
    & $\phi_1$ $(1, 1)_{Y}$ & $\phi_2$ $(3,2)_{7/6-Y}$ & $\psi_1$ $(3, 2)_{1/6-Y}$& $\psi_2$ $(\bar{3}, 1)_{-2/3+Y}$ &&&\\
    \hline
    $1$ & $\phi \ell\cdot \ell$ & $\phi\cdot \ell\bar{d}$ & $h\cdot\psi\bar{d}$ & $h^{\dagger}q\psi$ & No & No & Yes\\
    \hline
    $0$ & N/A & $\phi\cdot \ell\bar{u}$, $\phi^{\dagger}q\bar{e}$ & $h\cdot\psi\bar{u}$, $h^{\dagger}\psi\bar{d}$ & $q\cdot h\psi$ & Yes, $(q\bar{e})\cdot(\ell\bar{u})$ & Yes & No\\
    \hline
    $-1$ & $\phi^{\dagger} \ell\cdot \ell$ & N/A & $h^{\dagger}\psi\bar{u}$  & N/A & No & No & No\\
    \hline
    $2$ & $\phi^{\dagger}\bar{e}\bar{e}$ & N/A & N/A & N/A & No & No & Yes\\
    \hline
    $-2$ & $\phi\bar{e}\bar{e}$ & N/A & N/A & N/A & No & No & No\\
    \hline
\end{tabular}
\end{center}

\begin{center}
\begin{tabular}{ C{1cm} | C{2.7cm} | C{2.7cm} | C{2.7cm} | C{2.7cm} | c | c | c }
    \hline
    \multicolumn{8}{c}{$(q\bar{e})\cdot(\bar{u}\ell)$ Row 2}\\
    \hline
    \multirow{2}{*}{$Y$} & \multicolumn{4}{c|}{SM coupling} & \multirow{2}{*}{EDM at 1-loop?} & \multirow{2}{*}{Tree-level $C_S$?} & \multirow{2}{*}{Proton decay?}\\
    \cline{2-5}
    & $\phi_1$ $(1, 2)_Y$ & $\phi_2$ $(3, 1)_{7/6-Y}$ & $\psi_1$ $(3, 1)_{1/6-Y}$ & $\psi_2$ $(\bar{3}, 2)_{-2/3+Y}$ &&&\\
    \hline
    $-1/2$ & $\phi\cdot q\bar{d}$, $\phi\cdot \ell\bar{e}$, $\phi^{\dagger} q\bar{u}$ & N/A & $q\cdot h\bar{\psi}$ & $h^{\dagger}\bar{\psi}\bar{u}$ & No & Yes & No\\
    \hline
    $1/2$ & $\phi\cdot q\bar{u}$, $\phi^{\dagger} q \bar{d}$, $\phi^{\dagger}\ell\bar{e}$ & N/A & $h^{\dagger}q\bar{\psi}$ & $h\cdot\bar{\psi}\bar{u}$, $h^{\dagger}\bar{\psi}\bar{d}$ & No & Yes & No\\
    \hline
    $3/2$ & N/A & $\phi\bar{u}\bar{e}$, $\phi^{\dagger} q\cdot \ell$ & N/A & $h\cdot\bar{\psi}\bar{d}$ & Yes, $(q\cdot\ell)(\bar{u}\bar{e})$ & Yes & Yes\\
    \hline
    $5/2$ & N/A & $\phi\bar{d}\bar{e}$ & N/A & N/A & No & No & Yes\\
    \hline
\end{tabular}
\end{center}

\begin{center}
\begin{tabular}{ C{1cm} | C{2.7cm} | C{2.7cm} | C{2.7cm} | C{2.7cm} | c | c | c }
    \hline
    \multicolumn{8}{c}{$(q\bar{e})\cdot(\bar{u}\ell)$ Row 3}\\
    \hline
    \multirow{2}{*}{$Y$} & \multicolumn{4}{c|}{SM coupling} & \multirow{2}{*}{EDM at 1-loop?} & \multirow{2}{*}{Tree-level $C_S$?} & \multirow{2}{*}{Proton decay?}\\
    \cline{2-5}
    & $\phi_1$ $(3,2)_Y$ & $\phi_2$ $(1, 1)_{7/6-Y}$ & $\psi_1$ $(1, 1)_{1/6-Y}$ & $\psi_2$ $(1, 2)_{-2/3+Y}$ &&&\\
    \hline
    $1/6$ & $\phi\cdot \ell\bar{d}$ & $\phi \ell\cdot \ell$ & N/A & $h^{\dagger}\psi\bar{e}$ & No & No\\
    \hline
    $7/6$ & $\phi\cdot \ell\bar{u}$, $\phi^{\dagger}q\bar{e}$ & N/A & $h^{\dagger}\ell\bar{\psi}$ & $h^{\dagger}\bar{\psi}\bar{e}$ & Yes, $(q\bar{e})\cdot(\ell\bar{u})$ & Yes & Yes\\
    \hline
    $-5/6$ & N/A & $\phi^{\dagger}\bar{e}\bar{e}$ & $h^{\dagger}\ell\psi$ & $h\cdot\psi\bar{e}$ & No & No & Yes\\
    \hline
    $13/6$ & N/A & $\phi^{\dagger} \ell\cdot \ell$ & N/A & $h\cdot\bar{\psi}\bar{e}$ & No & No & No\\
    \hline
    $19/6$ & N/A & $\phi\bar{e}\bar{e}$ & N/A & N/A & No & No & No\\
    \hline
\end{tabular}
\end{center}

\begin{center}
\begin{tabular}{ C{1cm} | C{2.7cm} | C{2.7cm} | C{2.7cm} | C{2.7cm} | c | c | c }
    \hline
    \multicolumn{8}{c}{$(q\bar{e})\cdot(\bar{u}\ell)$ Row 4}\\
    \hline
    \multirow{2}{*}{$Y$} & \multicolumn{4}{c|}{SM coupling} & \multirow{2}{*}{EDM at 1-loop?} & \multirow{2}{*}{Tree-level $C_S$?} & \multirow{2}{*}{Proton decay?}\\
    \cline{2-5}
    & $\phi_1$ $(3, 1)_Y$ & $\phi_2$ $(1, 2)_{7/6-Y}$ & $\psi_1$ $(1, 2)_{1/6-Y}$ & $\psi_2$ $(1, 1)_{-2/3+Y}$ &&&\\
    \hline
    $-1/3$ & $\phi\bar{u}\bar{e}$, $\phi^{\dagger} q\cdot \ell$ & N/A & $h^{\dagger}\bar{\psi}\bar{e}$ & $h^{\dagger}\ell\bar{\psi}$ & Yes, $(q\cdot\ell)(\bar{u}\bar{e})$ & Yes & Yes\\
    \hline
    $-4/3$ & $\phi\bar{d}\bar{e}$ & N/A & $h\cdot\bar{\psi}\bar{e}$ & N/A & No & No & Yes\\
    \hline
    $2/3$ & N/A & $\phi\cdot q\bar{u}$, $\phi^{\dagger}q\bar{d}$, $\phi^{\dagger}\ell\bar{e}$ & $h^{\dagger}\psi\bar{e}$ & N/A & No & Yes & No\\
    \hline
    $5/3$ & N/A & $\phi\cdot q\bar{d}$, $\phi\cdot \ell\bar{e}$, $\phi^{\dagger} q\bar{u}$ & $h\cdot\psi\bar{e}$ & $h^{\dagger}\ell\psi$ & No & Yes & No\\
    \hline
\end{tabular}
\end{center}

\end{landscape}

\newpage

\bibliography{ref}
\bibliographystyle{utphys}

\end{document}